\documentclass[journal]{IEEEtran}

\usepackage{cite}
\usepackage{amsmath}
\usepackage[T1]{fontenc}
\usepackage[latin9]{inputenc}
\usepackage{units}
\usepackage{textcomp}
\usepackage{stmaryrd}
\usepackage[pdftex]{graphicx}
\usepackage[caption=false,font=footnotesize]{subfig}
\usepackage[english]{babel}
\usepackage{stfloats}
\usepackage{fixltx2e}
\usepackage{array}
\usepackage{color}
\usepackage{multirow}
\usepackage{amssymb}
\usepackage{stackrel}
\usepackage{fancyhdr}

\begin{document}
\bibliographystyle{unsrt}

\title{Dense Dual-Attention Network for Light Field Image Super-Resolution}

\author{Yu~Mo,
        Yingqian~Wang,
        Chao~Xiao,
        Jungang~Yang,
        Wei~An

\thanks{This work was partially supported in part by the National Natural Science Foundation of China (Nos. 61921001, 61972435, 61401474). Corresponding author: Jungang~Yang (e-mail: yangjungang@nudt.edu.cn).}

\thanks{Y.~Mo, Y.~Wang, C.~Xiao, J.~Yang and W.~An are with the College of Electronic Science and Technology, National University of Defense Technology, Changsha, China. (e-mail: moyunudt@163.com, wangyingqian16@nudt.edu.cn, xiaochao12@nudt.edu.cn, yangjungang@nudt.edu.cn, anwei@nudt.edu.cn).}}
\maketitle
\thispagestyle{fancy}
\fancyhead{}
\lhead{}
\lfoot{"Copyright © 20xx IEEE. Personal use of this material is permitted. However, permission to use this material for any other purposes must be obtained from the IEEE by sending an email to pubs-permissions@ieee.org."}
\cfoot{}
\rfoot{}

\begin{abstract}
Light field (LF) images can be used to improve the performance of image super-resolution (SR) because both angular and spatial information is available. It is challenging to incorporate distinctive information from different views for LF image SR. Moreover, the long-term information from the previous layers can be weakened as the depth of network increases. In this paper, we propose a dense dual-attention network for LF image SR. Specifically, we design a view attention module to adaptively capture discriminative features across different views and a channel attention module to selectively focus on informative information across all channels. These two modules are fed to two branches and stacked separately in a chain structure for adaptive fusion of hierarchical features and distillation of valid information. Meanwhile, a dense connection is used to fully exploit multi-level information. Extensive experiments demonstrate that our dense dual-attention mechanism can capture informative information across views and channels to improve SR performance. Comparative results show the advantage of our method over state-of-the-art methods on public datasets.

\end{abstract}


\section{Introduction}

With the commercial light field (LF) cameras (e.g., Lytro and RayTrix) being commonly used in the market, 4D LF images have many applications such as post-capture refocusing \cite{Wang2018, ng2005light}, depth inference \cite{zhang2016light, shin2018epinet, peng2020zero, mishiba2020fast}, 3D reconstruction \cite{kim2013scene, peng2017lf}, and de-occlusion \cite{DeOccNet}. Although 4D LF images provide both spatial and angular information, the low spatial resolution is an essential problem in exploiting their advantage. Consequently, image super-resolution (SR) algorithms have been widely investigated to enhance the spatial resolution of LF images.

Different from single image SR (SISR) \cite{RCAN, hu2019channel} and video SR \cite{yi2020progressive, SOFVSR20}, LF image SR can incorporate the structural prior among sub-aperture images (SAIs) to improve the performance. Most traditional LF image SR methods \cite{rossi2017graph, wanner2014variational, ghassab2019light, LFBM5D, farrugia2017super} use disparity information as correlations to enhance the spatial resolution. These methods explicitly register the SAIs using the estimated disparities. However, the performance of these methods heavily depends on the accuracy of disparity estimation. Although many algorithms have been proposed to estimate disparities among LF images, it is difficult to obtain accurate disparities due to the low spatial resolution of LF images.

Recently, deep learning-based methods \cite{LFNet, LFCNN2017, cheng2019light, resLF, meng2020high, LF-InterNet, ATO} have been successfully applied to LF image SR, and achieved promising performance. Zhang et al. \cite{resLF} decoded the LF structural information into four stacked SAIs along different angular directions, and super-resolved the center image using a multi-branch residual network. Wang et al. \cite{LF-InterNet} proposed an LF-InterNet to decouple and interact spatial and angular information for LF image SR. Meng et al. \cite{meng2020high} proposed a 4D convolutional network to handle the LF structural information, which fully exploits the LF information from all adjacent SAIs. Jin et al. \cite{ATO} proposed an all-to-one (ATO) framework to learn the structural information and proposed a structural consistency regularization approach. Due to the occlusions and non-Lambertian reflections in LFs, the information in different views and different channels is of different importance. Existing learning-based LF image SR methods \cite{resLF, LF-InterNet, ATO, wang2020light} treat the view-wise and channel-wise features equally for LF image SR, and cannot discriminatively use the informative information in an LF for further performance improvement.

To handle these issues, in this paper, we propose a dense dual-attention network (namely, DDAN) for LF image SR. Our method can effectively extract discriminative features across different views and channels. Given an LF image, an atrous spatial pyramid pooling (ASPP) module cascaded by two residual blocks is first used for initial feature extraction. These features are then fed to the dual-attention module to generate different weights. The dual-attention module consists a view attention branch and a channel attention branch whose basic modules are residual view attention block (RVAB) and residual channel attention block (RCAB), respectively. Specifically, a dense skip connection is used in each branch to fully exploit multi-level features and preserve important information. Afterwards, a fusion module is designed to aggregate features obtained by these two branches. Finally, an upscaling module is used to generate the SR result. Ablation studies are conducted on both synthetic and real-world datasets to demonstrate the effectiveness of our modules. Comparative experiments are further conducted to demonstrate the superior performance of our network as compared to state-of-the-art single image and LF image SR methods. The main contributions of our network are summarized as follows:
\begin{itemize}
\item[$\bullet$] We design an RVAB using view-wise statistics to adaptively rescale the feature maps across views, and design an RCAB using channel-wise statistics to model the interdependencies among channels.
\item[$\bullet$] A dual-attention module is constructed by respectively cascading a set of RVABs and RCABs with dense connections. Our dual-attention module can extract discriminative features across different views and channels. It is demonstrated that informative features are efficiently extracted by our proposed modules to improve SR performance.
\item[$\bullet$] We develop a DDAN for accurate LF image SR. Our network achieves state-of-the-art performance as compared to recent single image and LF image SR methods.
\end{itemize}

The rest of this paper is organized as follows: Section II introduces the related literature. In Section III, we describe the architecture of our network in details. Experimental results on synthetic and real-world datasets are presented in Section IV. Finally, conclusions are drawn in Section V.

\section{Related Works}

\subsection{Single Image SR}

The aim of single image SR (SISR) is to obtain a high resolution (HR) image from its low resolution (LR) version. In order to achieve this goal, many deep learning-based algorithms have been proposed. Among them, Dong et al. \cite{SRCNN2014, SRCNN2015} first used a convolutional neural network (CNN) to enhance the resolution of an image. They proposed a three-layer convolutional network named SRCNN to predict HR image and achieved superior performance over traditional methods. Afterwards, many deeper and wider network structures \cite{VDSR, 2016Deeply, li2020mdcn} were designed to improve the performance of SISR. Lim et al. \cite{EDSR} stacked modified residual blocks to construct a deep SR network named EDSR which won the champion of the NTIRE 2017 SISR challenge \cite{NTIRE2017}. Ma et al. \cite{ma2019image} aggregated dense hierarchical features progressively in a tree structure to improve SISR performance. Liu et al. \cite{liu2020residual} aggregated the local residual features to extract more powerful features for SISR. Zhang et al. \cite{RDN} proposed a residual dense network (RDN) with skip connections to fully use hierarchical feature representations. Lai et al. \cite{2017Deep} developed a pyramidal framework to construct a deep laplacian pyramid super-resolution network with three sub-networks. The input LR image was added to the output of each sub-network to obtain SR results. Recently, several comprehensive reviews on SISR \cite{anwar2020deep, wang2020survey, yang2019deep} have been published.

\subsection{LF Image SR}

As multiple views are captured by an LF camera, the complementary information can be used to enhance the spatial resolution. Existing methods for LF image SR can be divided into two categories: optimization-based and learning-based methods.

Optimization-based methods require accurate structure information of the scene as priors and use sub-pixel information to enhance the resolution of LF images. Bishop et al. \cite{bishop2011light} formulated a variantional Bayesian framework for LF image SR based on the estimated scene disparity. Mitra and Veeraraghavan \cite{mitra2012light} used a Gaussian mixture model to achieve LF image SR with a patch-based approach. Wanner et al. \cite{wanner2012globally, wanner2014variational} exploited the structure tensor into a variational framework for LF image SR. Farrugia et al. \cite{farrugia2017super} developed an example-based algorithm on the patch-volumes. The LR-HR patch-volumes are decomposed into several subspaces, and a linear projection function is learned to super-resolve LF images. Rossi and Frossard \cite{GB} developed an LF image SR approach based on a graph regularizer, which enforced the LF geometric structure with a roughly estimated disparity. Alain et al. \cite{LFBM5D} employed an LFBM5D filter to achieve LF image SR based on SRBM3D \cite{BM3D}. However, limited by the accuracy of disparity and representation ability of handcrafted features, the performance of traditional methods is not competitive.

Inspired by the successful application of CNN in SISR, many learning-based methods have been developed to implicitly learn LF geometric structures to enhance the resolution of LF images. Yoon et al. \cite{LFCNN2015, LFCNN2017} proposed the pioneering method LFCNN to enhance the angular and spatial resolution simultaneously in which the spatial resolution of each SAI was enhanced by SRCNN \cite{SRCNN2014}. Wang et al. \cite{LFNet} modeled spatial relations between horizontally or vertically neighbouring views by using a bidirectional recurrent CNN and then super-resolved LF images with stacked generalization technique. Cheng et al. \cite{cheng2019light} combined the internal similarity and the external correlation in LFs to enhance the image resolution. Yeung et al. \cite{LFSSR} constructed an alternate spatial-angular convolution to extract LF structural information with a single forward pass. Zhang et al. \cite{resLF} employed four stacked SAIs along different angular directions into a multi-stream residual network (resLF) to super-resolve LF images. Wang et al. \cite{LF-InterNet} proposed an LF-InterNet to decouple the LF structural information into angular and spatial information for LF image SR. They further proposed to use deformable convolutions to address the disparity issue among LF images, and proposed an LF-DFnet \cite{wang2020light} for LF image SR. Meng et al. \cite{meng2020high} developed a 4D CNN to fully use the LF information from all adjacent SAIs. Jin et al. \cite{ATO} cascaded two modules (i.e. an all-to-one (ATO) module and a structural consistency regularization module) for LF image SR.

\begin{figure*}[t]
\vspace{-0.0cm}
\centering{
{\includegraphics[width=15cm]{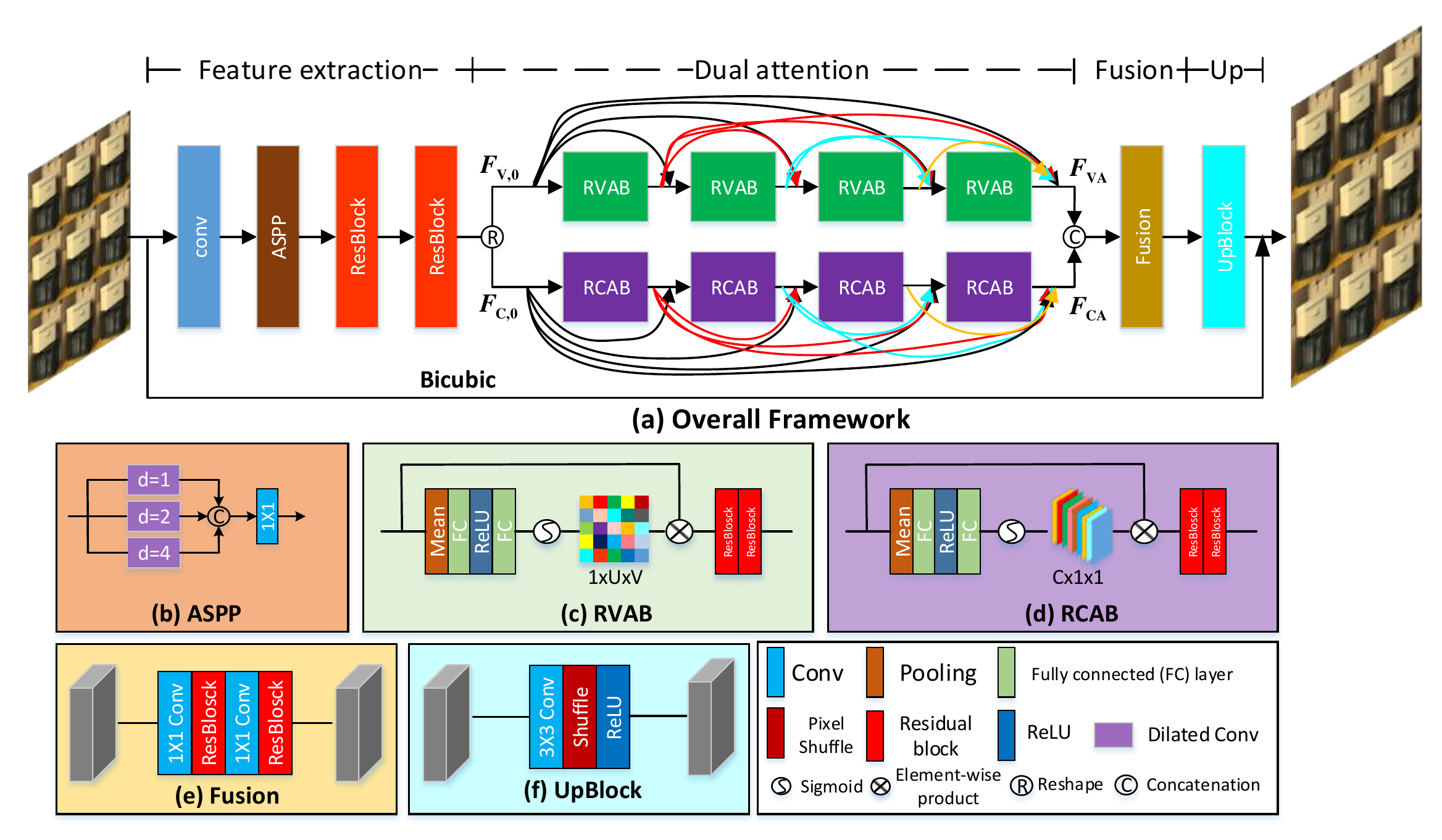}}
}
\caption{An overview of our network. (a) The overall architecture of the proposed DDAN network, which adopts attention mechanism and dense skip connection to improve SR performance. (b) The structure of the atrous spatial pyramid pooling module which is used to enlarge the receptive field. (c) The residual view attention block (RVAB), which adaptively rescales feature maps across views. (d) The residual channel attention block (RCAB), which model the interdependencies along channels. (e) The fusion module, which aggregates features obtained from dense dual-attention module. (f) The upscaling module, which generates SR images.}
\label{figure_network}
\end{figure*}

\subsection{Attention Mechanism in SR}

Attention mechanism plays an important role in computer vision \cite{woo2018cbam}. The aim of attention mechanism in CNN is to adaptively process features and focus on informative information. In recent years, some works have integrated attention modules into SR networks to improve their performance. Zhang et al. \cite{RCAN} proposed a residual channel attention network (RCAN) to recover more high-frequency information for SISR. In their method, channel attention mechanism was applied to model the interdependencies across feature channels. Hu et al. \cite{hu2019channel} proposed an SISR network including a channel attention module and a spatial attention module. These two attention modules are integrated by a gated fusion node to adaptively capture important information. Kim et al. \cite{kim2018ram} integrated spatial and channel attention with a novel fused scheme to build both inter-and-intra channel relationship. Dai et al. \cite{SAN} designed a second-order channel attention (SOCA) mechanism to model the inherent feature correlations in intermediate layers. In the area of stereo image SR, parallax attention mechanism \cite{PASSRnet, ying2020stereo, wang2020parallax, iPASSR} was widely used to incorporate cross-view information from stereo image pairs.

Inspired by the attention mechanism and considering that there are various information in different views of an LF, we develop a view-wise attention module to adaptively modulate the feature maps across views. Moreover, a channel-wise attention module is performed to model feature interdependencies across channels. It is demonstrated that the two attention modules can improve the discriminative learning ability of our network.

\section{Proposed Methods}

Following \cite{levoy1996light}, we denote a 4D LF as $\boldsymbol{L}(u,v,x,y)$, where $(u,v)$ represents the angular dimension and $(x,y)$ represents the spatial dimension. Following most existing SR methods \cite{resLF, ATO}, we only super-resolve the Y channel images which is converted from the RGB color space to YCbCr color space. As illustrated in Fig. \ref{figure_network}(a), our network consists of an initial feature extraction module, a dense dual-attention module, a fusion module, and an upscale module. In this section, we describe the proposed network in detail.

\subsection{Overview Architecture}

\begin{figure}
	\centering
	{\includegraphics[width=8.5cm]{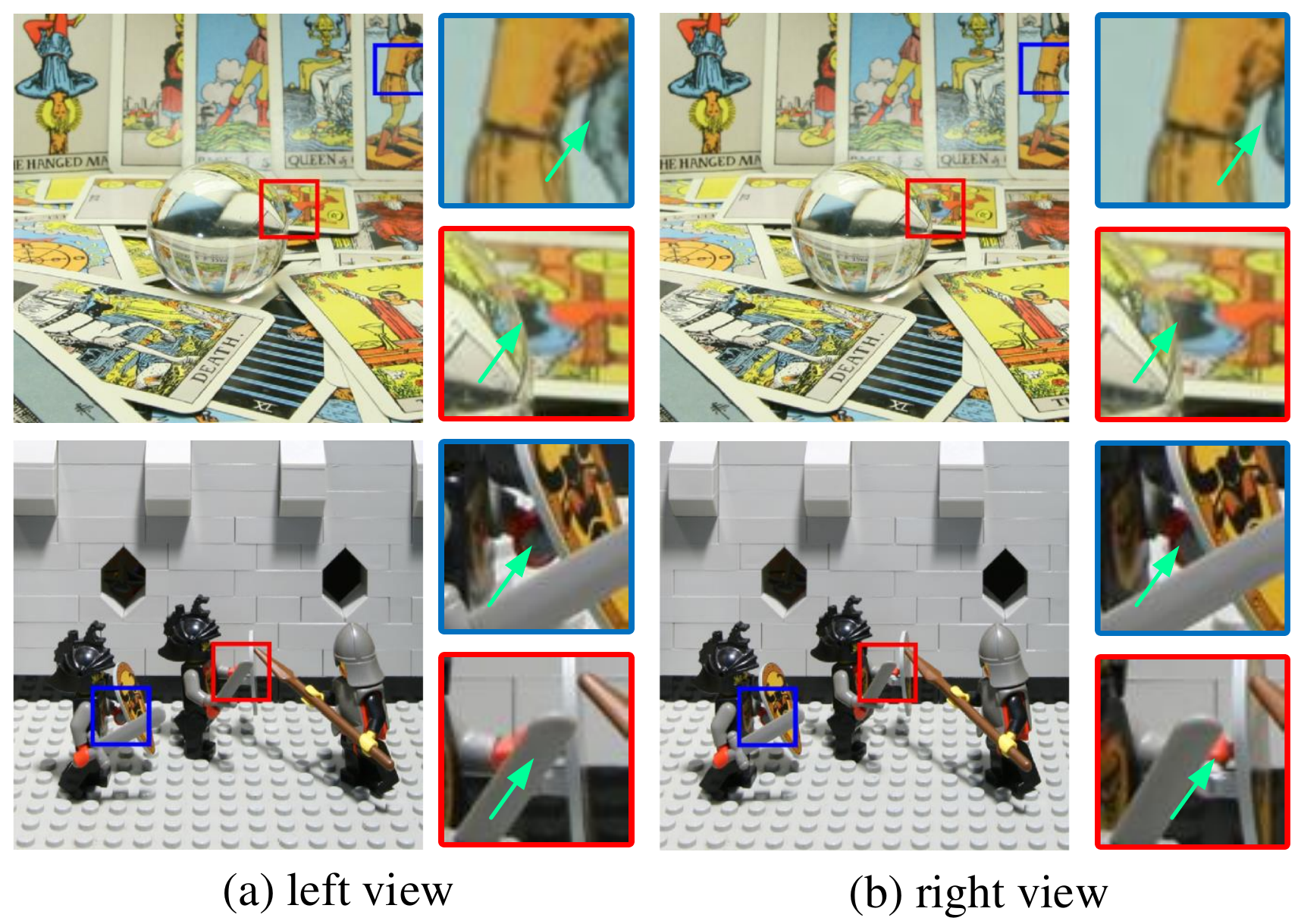}}
	\caption{The comparison of information in different views. The left view (a) and right view (b) are located at (5, 1) and (5, 9) of the $9\times 9$ LFs "Knights" and "Cards" respectively. One can clearly see that the different views contain different information especially near occlusion boundaries.}
	\label{figure_information}
\end{figure}

LF images are captured from the same scene with different angles. As shown in Fig. \ref{figure_information}, due to the occlusions and specularities, the details absent in one SAI may appear in another one, and such complementary information is helpful to predict the high frequency details of SR images. In order to capture distinctive information from various view images, we first extract shallow feature in each SAI with the sharing parameters (See Sec. III-B). Then two attention branches are designed to learn informative details from various views and channels by a sequence of stacked views and channel attention blocks (see Sec. III-C). To improve information flow across low-level and high-level layers within the two branches, dense connections \cite{RDN, tong2017SR, iPASSR} are used. The output of these two branches are fed into a fusion module and an upscale module to generate SR image of each SAI (see Sec. III-D).

Let $I_{LR}\in\mathbb{R}^{U\times V\times H\times W}$ denote an LR SAI array with $U\times V$ SAIs of resolution $H\times W$. Our network takes $I_{LR}\in\mathbb{R}^{U\times V\times H\times W}$ as its input and generates an HR SAI array of size $\mathbb{R}^{U\times V\times aH\times aW}$, where $a$ denotes the upsampling factor. The shallow features denoted as $F_{S}$ are extracted from each SAI individually by the shallow feature extraction module, i.e.,
\begin{equation}
F_{S}=H_{SF}(I_{LR}),
\end{equation}
where $F_{S}\in\mathbb{R}^{N\times C\times H\times W}$ represents the extracted shallow features. $C$ is the feature depth and $N=U\times V$ is the number of SAIs. ${H}_{SF}$ represents the shallow feature extractor. Then, we feed the shallow feature $F_{S}$ (termed as $F_{C,0}$ in the following text) into the channel attention branch and feed its reshaped feature $F_{V,0}\in\mathbb{R}^{C\times N\times H\times W}$ into the view attention branch. That is
\begin{equation}
F_{V,k}=H_{VA,k}(F_{V,0}+\ldots+F_{V,k-1}),
\end{equation}
\begin{equation}
F_{C,k}=H_{CA,k}(F_{C,0}+\ldots+F_{C,k-1}),
\end{equation}
where ${H}_{VA,k}$ and ${H}_{CA,k}$ represent the $k^{th}$ ($k\leq4$) view attention block and channel attention block, respectively. The output of the two branches are defined as
\begin{equation}
F_{VA}=F_{V,0}+F_{V,1}+F_{V,2}+F_{V,3}+F_{V,4},
\end{equation}
\begin{equation}
F_{CA}=F_{C,0}+F_{C,1}+F_{C,2}+F_{C,3}+F_{C,4},
\end{equation}
where the extracted deep feature maps ${F}_{VA}$ and ${f}_{CA}$ are fused by a fusion module.
\begin{equation}
F_{fu}=H_{Fusion}(F_{VA}, F_{CA}),
\end{equation}
where ${H}_{Fusion}$ denotes the fusion operation and ${F}_{fu}$ denotes the fused feature. After extracting informative features for the LF images, we introduce an up-sampling module to generate the SR images.
\begin{equation}
I_{SR}=U_{up}(F_{fu})+U_{bicubic}(I_{LR}),
\end{equation}
where ${U}_{up}$ is the upsampling operation with the efficient sub-pixel convolutional layer \cite{PixelShuffle} and ${U}_{bicubic}$ is the bicubic interpolation process.

\subsection{Shallow Feature Extraction}

Due to the occlusions, specularities and other factors, the information in different views is various. To extract discriminative features with rich context information, we use an atrous spatial pyramid pooling (ASPP) module to extract shallow features. It is demonstrated in \cite{PASSRnet, wang2020parallax} that the ASPP module is beneficial for extracting hierarchical features and can enlarge the receptive field.

As shown in Fig. \ref{figure_network}(b), we first put each SAI into a same $3\times 3$ convolution to extract initial features. Input features are fed to an ASPP to generate multi-scale features. Then two residual blocks are used for deep feature extraction. The ASPP module is constructed by three dilated convolutions parallelly with dilation rates of 1, 2, 4, respectively. Each residual block includes two $3\times 3$ convolution layers and a ReLU layer. The ASPP module can effectively improve the performance of SR, as demonstrated in Sec. IV-D.

\subsection{Dense Dual-Attention Module}

The features generated by the shallow feature extraction module contain many types of information across views and channels. A straightforward approach to integrate these information is to feed these features into a multi-layer network which deals with them equally. However, using the aforementioned approach will limit the representational ability of the network. To improve the SR performance and better recover the high-frequency details, it is important to consider the different contributions of these features. By utilizing the view-wise and channel-wise statistics among these features, we design a view attention branch to adaptively rescale the feature maps across views, and design a channel attention branch to model the interdependencies among channels. 

For the view attention branch, we operate on the second dimension $N$ of the input $F_{V,0}\in\mathbb{R}^{C\times N\times H\times W}$ to make fully use of view-specific characteristics among SAIs. Specifically, for each slice in dimension $C$, feature volume $\mathbb{R}^{N\times H\times W}$ is obtained from all SAIs with same weights. Due to the occlusions, specularities and other factors, different views contain different information. As a result, we propose to capture discriminative features in each feature volume using attention mechanism. Moreover, for the channel attention branch, we process the second dimension $C$ of the input $F_{C,0}\in\mathbb{R}^{N\times C\times H\times W}$ to model the interdependencies among channels. For each view feature $\mathbb{R}^{C\times H\times W}$, treating each slice of them $\mathbb{R}^{H\times W}$ equally hinders the representation capability of networks. So we use attention mechanism to selectively focus on discriminative information across all channels within the same view.

\begin{figure*}[t]
\vspace{-0.0cm}
\centering{
{\includegraphics[width=15cm]{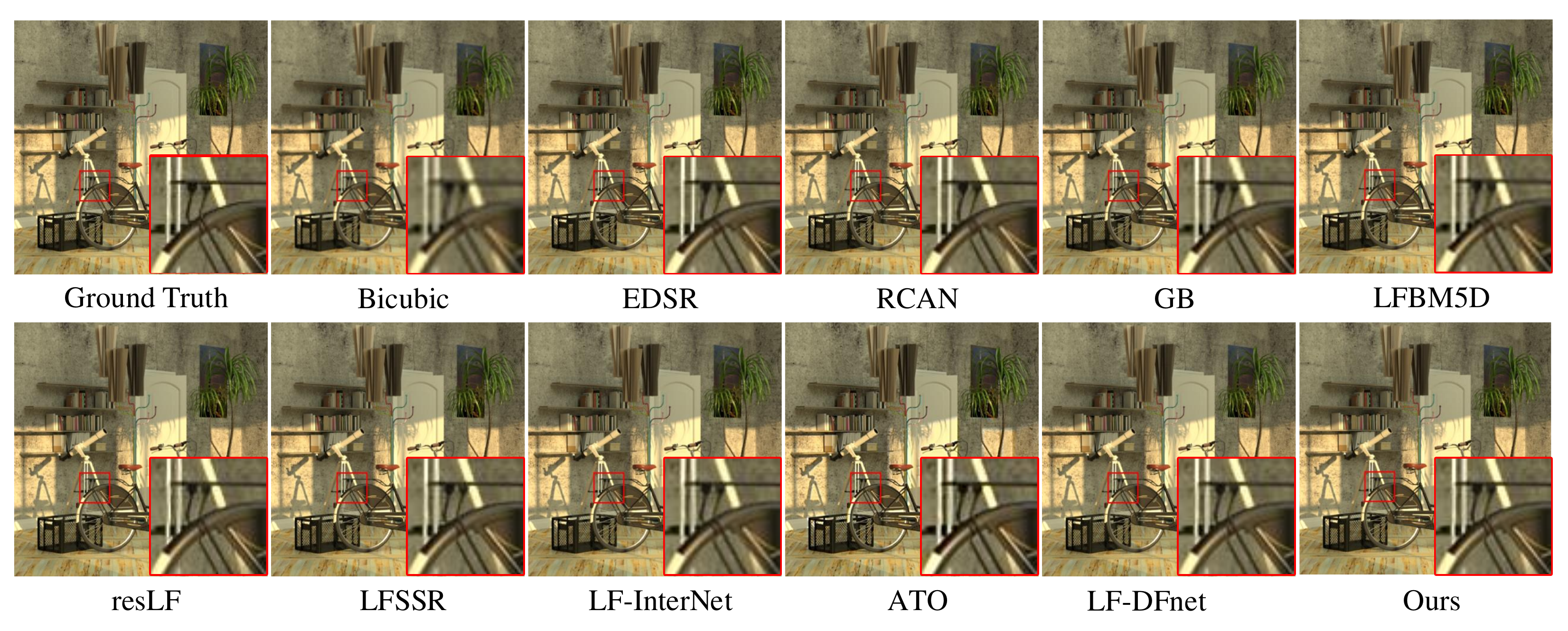}}}

\caption{Visual comparisons (reconstructed central SAIs) achieved by different methods on scene "Bicycle" for $2\times $ SR.}
\label{figure_2xSR_bicycle}
\end{figure*}

\begin{figure*}[t]
\vspace{-0.0cm}
\centering{
{\includegraphics[width=15cm]{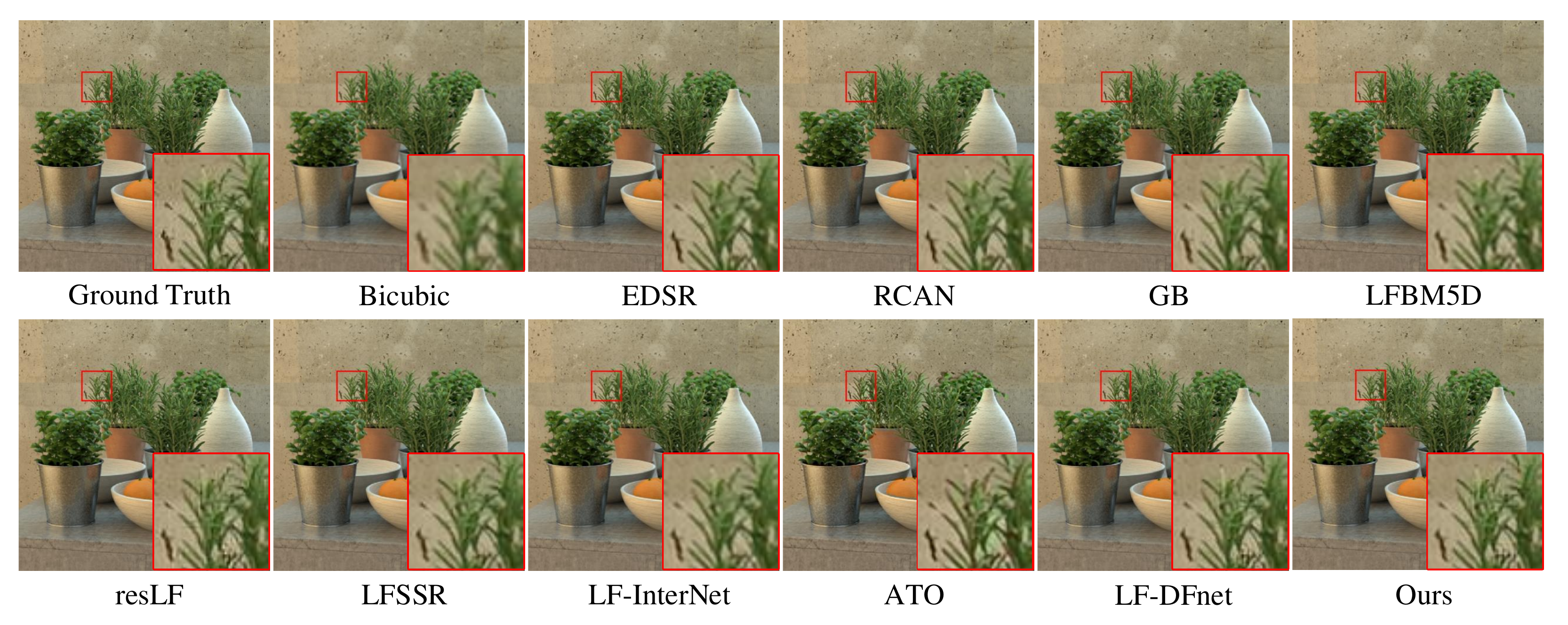}}}

\caption{Visual comparisons (reconstructed central SAIs) achieved by different methods on scene "Herbs" for $2\times $ SR.}
\label{figure_2xSR_herbs}
\end{figure*}

\textbf{View Attention Branch}.
Previous CNN-based LFSR methods treat LR view-wise features equally, which limits the full exploitation of the complementary information among SAIs. Due to the occlusions, specularities and other factors, adaptively selecting features from various views is helpful to boost the representation capability of the network. The structure of the view attention branch is constructed by cascading RVABs for four times. As shown in Fig. \ref{figure_network}(c), we stack an attention module with two residual blocks within each RVAB and apply dense skip connections among blocks to effectively integrate the low-level and high-level features. In each attention module, global average pooling is used to obtain statistics of features. Then we obtain attention maps with a gating mechanism which consists of two fully connected (FC) layers and a ReLU layer. To recalibrate the features across all views, the attention map is normalized by a sigmoid function with a range from 0 to 1. We denote $X=[x_{1},x_{2},\cdots,x_{N}]$ as the input of the view attention branch, which consists of $N$ feature maps of size $H\times W$. The view-wise statistic $z\in\mathbb{R}^{N\times 1\times 1}$ can be obtained by the average pooling which is operated on individual view features over spatial dimensions $H\times W$. The $n^{th}$ element of $z$ is determined by
\begin{equation}
z_{n}=\frac{1}{H\times W}\stackrel[i=1]{H}{\sum}\stackrel[j=1]{W}{\sum}x_{n}(i,j),
\end{equation}
where ${x}_{n}(i,j)$ is the value of the view feature ${x}_{n}$ at position $(i,j)$.

\begin{figure*}[t]
\vspace{-0.0cm}
\centering{
{\includegraphics[width=15cm]{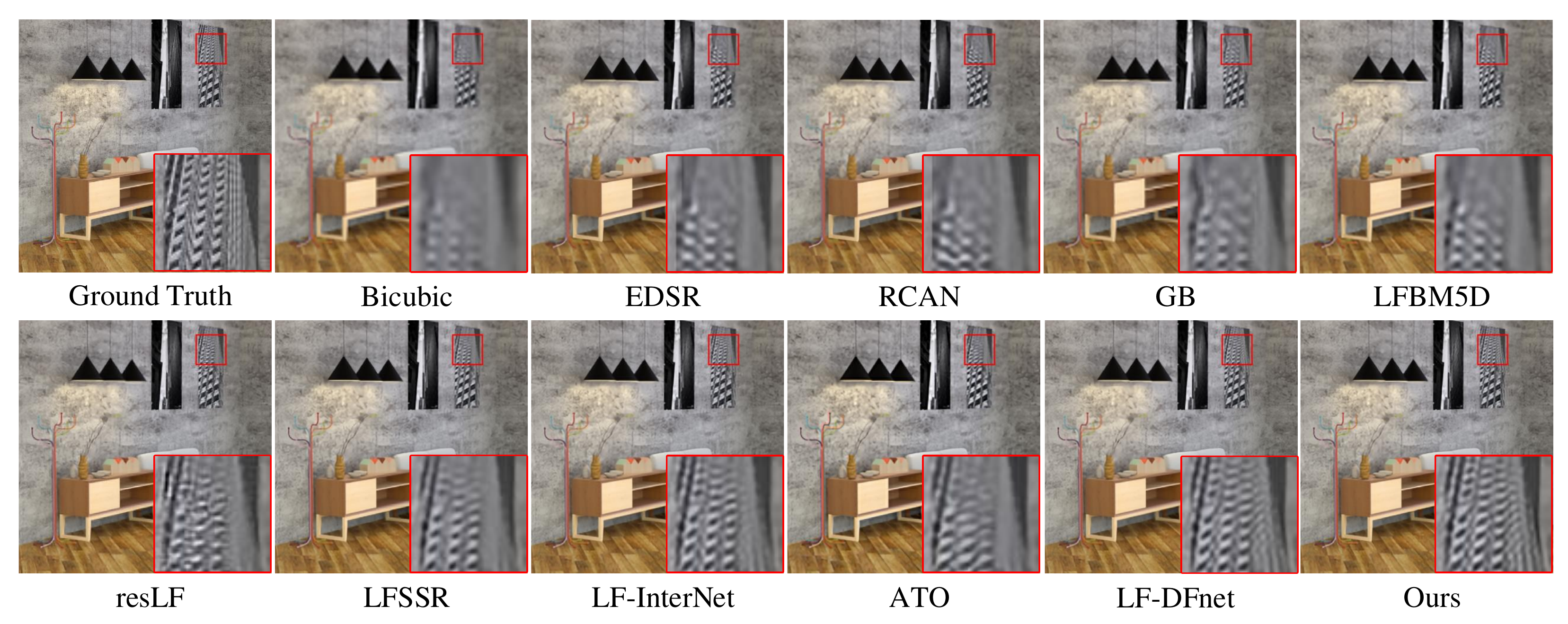}}}

\caption{Visual comparisons (reconstructed central SAIs) achieved by different methods on scene "Bedroom" for $4\times $ SR.}
\label{figure_4xSR_bedroom}
\end{figure*}

\begin{figure*}[t]
\vspace{-0.0cm}
\centering{
{\includegraphics[width=15cm]{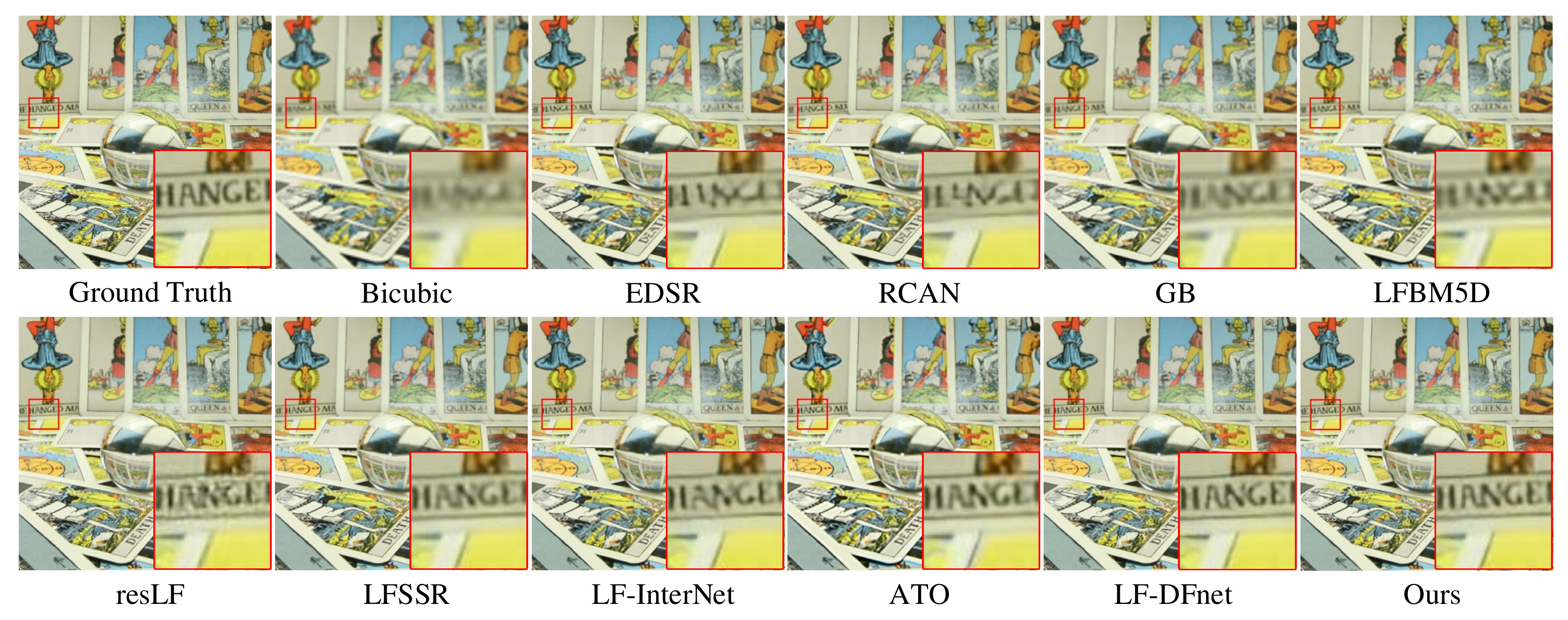}}}

\caption{Visual comparisons (reconstructed central SAIs) achieved by different methods on scene "Cards" for $4\times $ SR.}
\label{figure_4xSR_cards}
\end{figure*}

\begin{figure*}[t]
\vspace{-0.0cm}
\centering{
{\includegraphics[width=15cm]{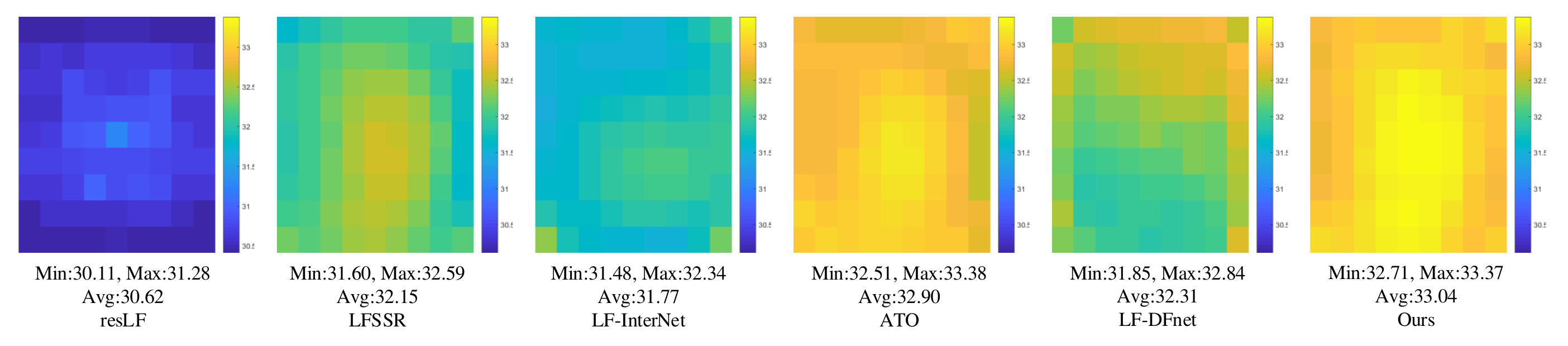}}
}
\caption{Comparison of the PSNR distribution on light field scene "Rusty Fency" for $2\times $ SR. Each view is visualized with a grid and different colors represent different PSNR values. The color range is set to $(30.11, 33.38)$. }
\label{figure_std}
\end{figure*}

As discussed in \cite{RCAN}, a gating mechanism is needed to assign various attentions to different view features. We utilize an FC-ReLU-FC\footnote{In our implementation, FC is achieved by using a $1\times 1$ convolution.} layer with a sigmoid activation as the gating mechanism to aggregate information across all views, which is defined by
\begin{equation}
w=\sigma(W_{2}\ast\delta(W_{1}\ast z),
\end{equation}
where $\sigma(\cdot)$ and $\delta(\cdot)$ denote the sigmoid and ReLU functions, respectively. $w\in\mathbb{R}^{N\times1\times1}$ is the view-wise attention weights and $\ast$ is the convolution operation. $W_{1}\in\mathbb{R}^{N\times\frac{N}{\xi}\times 1\times1}$ is the weight in the first convolutional layer which is used to downscale the number of features with reduction ratio $\xi$. Next, these features are activated by ReLU and the number of them are increased with ratio $\xi$ (the effect of the reduction ratio is investigated in Sec. IV-D) by another convolutional layer whose weight is $W_{2}\in\mathbb{R}^{\frac{N}{\xi}\times N\times1\times1}$. Finally, we rescale the input features with the attention map $w$ according to
\begin{equation}
\stackrel{\wedge}{x_{n}}=w_{n}\cdot x_{n},
\end{equation}
where $w_{n}$ and $x_{n}$ denote the scaling weight and feature map of the $n^{th}$ view, respectively.

With the RVAB and dense skip connections, the input features are adaptively modulated according to their statistics. As a result, the view attention branch can preserve view-specific characteristics, which is beneficial to improve the discriminative ability of the network on different views. The effectiveness of the view attention module is demonstrated in Sec. IV-D.

\textbf{Channel Attention Branch}.
The aim of the channel attention branch is to selectively focus on discriminative information across all channels of a view. The structure of the channel attention branch consists of four RCABs which are connected with dense skip connections. As shown in Fig. \ref{figure_network}(d), we integrate a channel attention module with two residual blocks to build our RCAB. The channel-wise statistic $s\in\mathbb{R}^{C\times1\times1}$ can be generated by performing global average pooling on the input features. Similar to view attention block, an FC-ReLU-FC structure followed by a sigmoid activation is exploited to generate a channel attention map $m$, i.e.
\begin{equation}
m=\sigma(M_{2}\ast\delta(M_{1}\ast s),
\end{equation}
where the meanings of the notations $\sigma(\cdot)$ and $\delta(\cdot)$ are the same as those in Eq. (9). The first layer with parameters of $M_{1}\in\mathbb{R}^{C\times\frac{C}{\theta}\times 1\times1}$ is used to downscale the number of features with a reduction ratio $\theta$ (the effect of the reduction ratio is investigated in Sec. IV-D). Further, the result is combined into a ReLU and then upscaled with ratio $\theta$ by another convolutional layer whose weight is $M_{2}\in\mathbb{R}^{\frac{C}{\theta}\times C\times1\times1}$. Finally, the rescaled features are obtained by
\begin{equation}
\stackrel{\wedge}{y_{c}}=m_{c}\cdot y_{c},
\end{equation}
where $m_{c}$ and $y_{c}$ denote the scaling weight and feature map in the $c^{th}$ channel, respectively.

Due to the employment of the channel attention module, the channel-wise features are adaptively rescaled at multiple levels to emphasize important features and suppress unnecessary ones. We show the effectiveness of the channel attention module in Sec. IV-D.

\subsection{Fusion and Upscaling module}

The fusion module is introduced to fuse the view features $F_{VA}\in\mathbb{R}^{C\times N\times H\times W}$ and channel features $F_{CA}\in\mathbb{R}^{N\times C\times H\times W}$. We reshape them to the same size and then concatenate them. To obtain highly informative features, the concatenated features $F\in\mathbb{R}^{N\times 2C\times H\times W}$ are fused by a fusion module, which consists of two $1\times 1$ convolution layers and two residual blocks (as shown in Fig. \ref{figure_network}(e)). Inspired by the efficient sub-pixel convolutional layer \cite{PixelShuffle}, a pixel-shuffle layer followed by a $3\times 3$ convolution layer is used to upscale the fused features to the target resolution $aH\times aW$ (Fig. \ref{figure_network}(f)).

The proposed network super-resolves each SAI and is optimized by minimizing the difference between the SR image ${I}_{SR}$ and the corresponding ground-truth image ${I}_{HR}$. ${L}_{1}$ loss function is applied to measure the difference.
\begin{equation}
L=\frac{1}{N}\sum_{k=1}^{N}||I_{SR}^{k}-I_{HR}^{k}||_{1}.
\end{equation}

\begin{figure*}[t]
\vspace{-0.0cm}
\centering{
{\includegraphics[width=15cm]{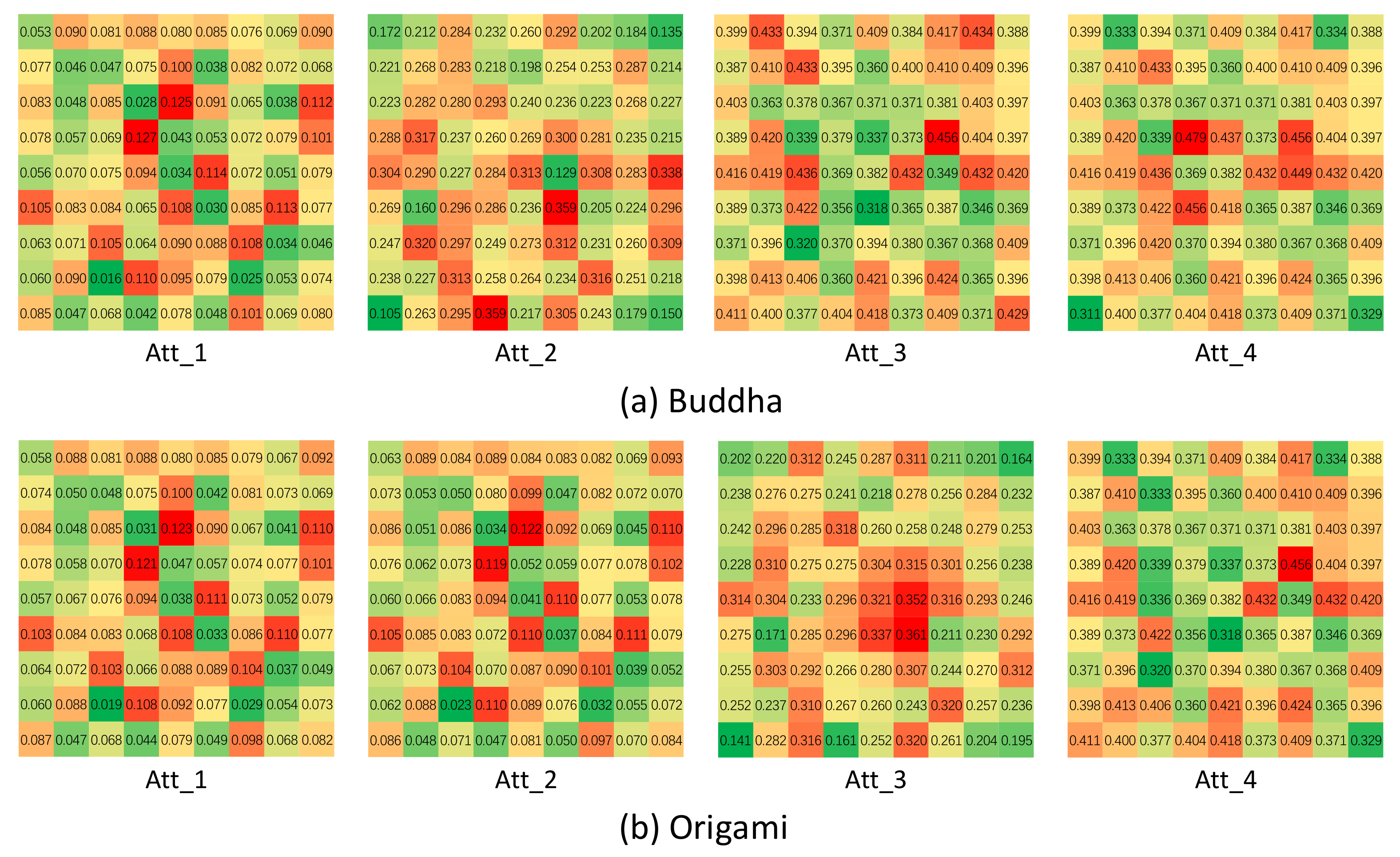}}}

\caption{Visualization of the attention weights for two different LF images. Here, we compare each attention module in the view attention branch. $Att\_1$, $Att\_2$, $Att\_3$ and $Att\_4$ represent the first, second, third and fourth attention module, respectively.}
\label{figure_att_weights}
\end{figure*}

\section{Experiments}

\begin{table}
\renewcommand\arraystretch{1.2}
\centering
\caption{The details of datasets used in our experiments.}

\begin{tabular}{|c|cc|ccc|}
\hline
 & \multicolumn{2}{c|}{Synthetic} & \multicolumn{3}{c|}{Real-world}\tabularnewline
\hline
\multirow{2}{*}{Datasets} & HCInew & HCIold & EPFL & INRIA & STFgantry\tabularnewline
 & \cite{honauer2016dataset} & \cite{wanner2013datasets} & \cite{EPFL} & \cite{INRIA} & \cite{STFgantry}\tabularnewline
\hline
Training & 20 & 10 & 70 & 35 & 9\tabularnewline
Test & 4 & 2 & 10 & 5 & 2\tabularnewline
Disparity & {[}-4,4{]} & {[}-3,3{]} & {[}-1,1{]} & {[}-1,1{]} & {[}-7,7{]}\tabularnewline
\hline
\end{tabular}
\label{table_datasets}
\end{table}

In this section, we first introduce the datasets and implementation details, and then compare our network to several state-of-the-art single image and LF image SR methods. Finally, we conduct ablation study to investigate our network.

\subsection{Datasets}

As listed in Table \ref{table_datasets}, the synthetic LF datasets from HCInew \cite{honauer2016dataset} and HCIold \cite{wanner2013datasets}, and real-world LF datasets from EPFL \cite{EPFL}, INRIA \cite{INRIA}, STFgantry \cite{STFgantry} were used in experiments. Specifically, 144 LF images including 30 synthetic images and 114 real-world images were used for training, and 23 LF images containing 6 synthetic scenes and 17 real-world scenes were used for test. To be specific, different datasets contain different scenes with various disparity range, ensuring the diversity of datasets. All LFs in these datasets were processed with angular resolutions of $5\times 5$ and $9\times 9$. Each SAI was cropped into patches of size $64\times 64$ with a stride of 32. These patches were spatially $\times 2$ and $\times 4$ downsampled using the bicubic interpolation approach to generate input LR patches. Further, these patches were randomly flipped and 90-degree rotated along spatial and angular directions simultaneously for data augmentation \cite{wang2020light}.

\subsection{Implementation Details}

In our network, $C$ was set to 32, $N$ was 25 for $5\times 5$ input LFs and 81 for $9\times 9$ input LFs. The network weights was initialized using the Kaiming method \cite{KaimingInit} and optimized using the Adam method \cite{Adam} with ${\beta}_{1}=0.9$ and ${\beta}_{2}=0.999$. The reduction ratios $\xi$ and $\theta$ in attention blocks are set to 2. Our network was implemented in PyTorch on a PC with an NVidia RTX 2080Ti GPU. The batch size was set to 30. The learning rate was initially set to $5\times10^{-4}$, and reduced to half after every 20 epochs. The training was stopped after 80 epochs since more epochs can not further introduce performance improvements.

\begin{table*}
\renewcommand\arraystretch{1.2}
\caption{Quantitative comparisons of state-of-the-art methods on $2\times $ and $4\times $ LF image SR. The PSNR/SSIM are the average value of all the scenes of a dataset. The results presented on $5\times 5$ and $9\times 9$ LFs. The best results are in \textcolor{red}{red} and the second best results are in \textcolor{blue}{blue}.}
\centering
\resizebox{\textwidth}{80pt}{
\begin{tabular}{|c|c|c|c|c|c|c|ccccc|}
\hline
\multirow{2}{*}{Method} & \multirow{2}{*}{Scale} & \multicolumn{5}{c|}{Dataset (Input 5x5)} & \multicolumn{5}{c|}{Dataset (Input 9x9)}\tabularnewline
\cline{3-12}
 &  & EPFL \cite{EPFL} & HCInew \cite{honauer2016dataset} & HCIold \cite{wanner2013datasets} & INRIA \cite{INRIA} & STFgantry \cite{STFgantry} & \multicolumn{1}{c|}{EPFL \cite{EPFL}} & \multicolumn{1}{c|}{HCInew \cite{honauer2016dataset}} & \multicolumn{1}{c|}{HCIold \cite{wanner2013datasets}} & \multicolumn{1}{c|}{INRIA \cite{INRIA}} & STFgantry \cite{STFgantry}\tabularnewline
\hline
Bicubic & \multirow{10}{*}{$2\times$} & 29.50/0.9350 & 31.69/0.9335 & 37.46/0.9776 & 31.10/0.9563 & 30.82/0.9473 & 29.41/0.9334 & 31.58/0.9325 & 37.32/0.9765 & 31.02/0.9551 & 30.71/0.9462\tabularnewline
\cline{1-1}
EDSR \cite{EDSR} &  & 33.09/0.9631 & 34.83/0.9594 & 41.01/0.9875 & 34.97/0.9765 & 36.29/0.9819 & 33.01/0.9620 & 34.77/0.9591 & 40.94/0.9871 & 34.92/0.9759 & 36.22/0.9814\tabularnewline
\cline{1-1}
RCAN \cite{RCAN} &  & 33.16/0.9635 & 34.98/0.9602 & 41.05/0.9875 & 35.01/0.9769 & 36.33/0.9825 & 33.25/0.9645 & 35.04/0.9611 & 41.02/0.9875 & 35.02/0.9771 & 36.31/0.9822\tabularnewline
\cline{1-1}
LFBM5D \cite{LFBM5D} &  & 31.15/0.9545 & 33.72/0.9548 & 39.62/0.9854 & 32.85/0.9695 & 33.55/0.9718 & 31.21/0.9549 & 33.78/0.9546 & 39.68/0.9858 & 32.93/0.9704 & 33.68/0.9714\tabularnewline
\cline{1-1}
GB \cite{GB} &  & 31.22/0.9591 & 35.25/0.9692 & 40.21/0.9879 & 32.76/0.9724 & 35.44/0.9835 & 31.28/0.9590 & 35.16/0.9688 & 40.12/0.9876 & 32.78/0.9726 & 35.36/0.9827\tabularnewline
\cline{1-1}
resLF \cite{resLF} &  & 32.75/0.9672 & 36.07/0.9715 & 42.61/0.9922 & 34.57/0.9784 & 36.89/0.9873 & 32.86/0.9681 & 36.11/0.9718 & 42.68/0.9924 & 34.58/0.9782 & 36.95/0.9977\tabularnewline
\cline{1-1}
LFSSR \cite{LFSSR} &  & 33.69/0.9748 & 36.86/0.9753 & 43.75/0.9939 & 35.27/0.9834 & 38.07/0.9902 & 34.64/0.9767 & 36.15/0.9674 & 43.64/0.9871 & 36.12/0.9829 & 37.18/0.9825\tabularnewline
\cline{1-1}
LF-InterNet \cite{LF-InterNet} &  & 34.14/0.9761 & 37.28/0.9769 & \textcolor{blue}{44.45}/0.9945 & 35.80/\textcolor{blue}{0.9846} & 38.72/0.9916 & 34.45/0.9752 & 35.59/0.9688 & 43.76/0.9877 & 36.13/0.9832 & 36.03/0.9844\tabularnewline
\cline{1-1}
ATO \cite{ATO} &  & 34.41/0.9762 & 37.01/0.9761 & 43.79/0.9940 & 36.21/0.9840 & 39.01/\textcolor{blue}{0.9921} & \textcolor{blue}{34.94}/\textcolor{red}{0.9789} & 36.90/0.9763 & 44.10/\textcolor{blue}{0.9941} & 36.57/0.9842 & 39.03/\textcolor{blue}{0.9923}\tabularnewline
\cline{1-1}
LF-DFnet \cite{wang2020light} &  & \textcolor{blue}{34.44}/\textcolor{blue}{0.9766} & \textcolor{blue}{37.44}/\textcolor{blue}{0.9786} & 44.23/\textcolor{red}{0.9947} & \textcolor{blue}{36.36}/0.9841 & \textcolor{red}{39.61}/\textcolor{red}{0.9935} & 34.74/0.9770 & \textcolor{blue}{37.02}/\textcolor{blue}{0.9768} & \textcolor{blue}{44.14}/0.9940 & \textcolor{blue}{36.65}/\textcolor{blue}{0.9844} & \textcolor{red}{39.43}/\textcolor{red}{0.9926}\tabularnewline
\cline{1-1}
Ours &  & \textcolor{red}{35.04}/\textcolor{red}{0.9774} & \textcolor{red}{37.68}/\textcolor{red}{0.9799} & \textcolor{red}{44.47}/\textcolor{blue}{0.9946} & \textcolor{red}{37.34}/\textcolor{red}{0.9855} & \textcolor{blue}{39.14}/0.9913 & \textcolor{red}{35.48}/\textcolor{blue}{0.9786} & \textcolor{red}{37.31}/\textcolor{red}{0.9780} & \textcolor{red}{44.41}/\textcolor{red}{0.9944} & \textcolor{red}{37.97}/\textcolor{red}{0.9850} & \textcolor{blue}{39.08}/\textcolor{red}{0.9926}\tabularnewline
\hline
Bicubic & \multirow{12}{*}{$4\times$} & 25.14/0.8311 & 27.61/0.8507 & 32.42/0.9335 & 26.82/0.8860 & 25.93/0.8431 & 25.10/0.8304 & 27.54/0.8502 & 32.36/0.9328 & 26.77/0.8853 & 25.90/0.8427\tabularnewline
\cline{1-1}
EDSR \cite{EDSR} &  & 27.84/0.8858 & 29.60/0.8874 & 35.18/0.9538 & 29.66/0.9259 & 28.70/0.9075 & 27.81/0.8857 & 29.54/0.8872 & 35.16/0.9533 & 29.62/0.9255 & 28.67/0.9071\tabularnewline
\cline{1-1}
RCAN \cite{RCAN} &  & 27.88/0.8863 & 29.63/0.8880 & 35.20/0.9540 & 29.76/0.9273 & 28.90/0.9110 & 27.94/0.8869 & 29.70/0.8891 & 35.22/0.9541 & 29.82/0.9277 & 28.96/.9115\tabularnewline
\cline{1-1}
LFBM5D \cite{LFBM5D} &  & 26.61/0.9689 & 29.13/0.8823 & 34.23/0.9510 & 28.49/0.9137 & 28.30/0.9002 & 26.68/0.8692 & 29.19/0.8825 & 34.28/0.9514 & 28.58/0.9141 & 28.39/0.9010\tabularnewline
\cline{1-1}
GB \cite{GB} &  & 26.02/0.8628 & 28.92/0.8842 & 33.74/0.9497 & 27.73/0.9085 & 28.11/0.9002 & 26.08/0.8630 & 28.95/0.8844 & 33.81/0.9451 & 27.79/0.9088 & 28.13/0.9015\tabularnewline
\cline{1-1}
resLF \cite{resLF} &  & 27.46/0.8899 & 29.92/0.9011 & 36.12/0.9651 & 29.64/0.9339 & 28.99/0.9214 & 27.52/0.8891 & 29.96/0.9013 & 36.21/0.9562 & 29.76/0.9351 & 29.11/0.9402\tabularnewline
\cline{1-1}
LFSSR \cite{LFSSR} &  & 28.27/0.9080 & 30.72/0.9124 & 36.70/0.9690 & 30.31/0.9446 & 30.15/0.9385 & 28.34/0.9085 & 30.12/0.9103 & 36.12/0.9645 & 30.34/0.9436 & 29.25/0.9391\tabularnewline
\cline{1-1}
LF-InterNet \cite{LF-InterNet} &  & 28.67/0.9143 & 30.98/0.9165 & 37.11/0.9715 & 30.64/0.9486 & 30.53/\textcolor{blue}{0.9426} & 28.62/0.9134 & 29.72/0.9084 & 36.44/0.9671 & 30.46/0.9442 & 29.13/0.9387\tabularnewline
\cline{1-1}
ATO \cite{ATO} &  & \textcolor{blue}{28.92}/\textcolor{blue}{0.9171} & 30.16/0.9101 & 36.21/0.9662 & \textcolor{blue}{31.05}/0.9464 & 29.89/0.9382 & \textcolor{blue}{29.13}/\textcolor{blue}{0.9178} & 30.39/0.9103 & 36.28/0.9667 & \textcolor{blue}{31.22}/\textcolor{blue}{0.9473} & 29.99/0.9384\tabularnewline
\cline{1-1}
LF-DFnet \cite{wang2020light} &  & 28.77/0.9165 & \textcolor{blue}{31.23}/\textcolor{blue}{0.9196} & \textcolor{blue}{37.32}/\textcolor{blue}{0.9718} & 30.83/\textcolor{blue}{0.9503} & \textcolor{red}{31.15}/\textcolor{red}{0.9494} & 29.07/0.9167 & \textcolor{blue}{30.92}/\textcolor{blue}{0.9133} & \textcolor{blue}{36.82}/\textcolor{blue}{0.9701} & 30.74/0.9472 & \textcolor{blue}{30.86}/\textcolor{blue}{0.9441}\tabularnewline
\cline{1-1}
Ours &  & \textcolor{red}{29.19}/\textcolor{red}{0.9178} & \textcolor{red}{31.60}/\textcolor{red}{0.9211} & \textcolor{red}{37.36}/\textcolor{red}{0.9722} & \textcolor{red}{31.44}/\textcolor{red}{0.9513} & \textcolor{blue}{30.72}/0.9419 & \textcolor{red}{29.75}/\textcolor{red}{0.9235} & \textcolor{red}{31.48}/\textcolor{red}{0.9216} & \textcolor{red}{36.97}/\textcolor{red}{0.9736} & \textcolor{red}{32.05}/\textcolor{red}{0.9543} & \textcolor{red}{31.04}/\textcolor{red}{0.9539}\tabularnewline
\hline
\end{tabular}}
\label{table_results}
\end{table*}

\subsection{Comparison to State-of-the-arts}
We compare the proposed method with state-of-the-art methods including two SISR methods i.e., EDSR \cite{EDSR}, RCAN \cite{RCAN}), and seven LFSR methods i.e., LFBM5D \cite{LFBM5D}, GB \cite{GB}, resLF \cite{resLF}, LFSSR \cite{LFSSR}, LF-InterNet \cite{LF-InterNet}, ATO \cite{ATO} and LF-DFnet \cite{wang2020light}. Bicubic interpolation was included as baselines. We have retrained all learning-based methods \cite{EDSR, RCAN, resLF, LFSSR, LF-InterNet, ATO, wang2020light} on the same training datasets as ours. The peak signal-to-noise ratio (PSNR) and the structural similarity index (SSIM) \cite{wang2004image} were used for quantitative evaluation.

\begin{table*}
	\renewcommand\arraystretch{1.6}
	\centering
	\caption{Comparisons of the number of parameters, FLOPs and running time (in seconds) for $2\times $ and $4\times $ LF spatial SR. FLOPs are calculated on $5\times 5\times 32\times 32$ and $9\times 9\times 32\times 32$ input features. Running time is calculated in super-resolving an LF. Here, we use PSNR value averaged over whole test datasets to represent their reconstruction accuracy.}
	\resizebox{\textwidth}{45pt}{
		\begin{tabular}{|c|c|c|c|c|c|c|c|c|c|c|c|c|c|c|c|c|c|c|}
			\hline 
			\multirow{2}{*}{Method} & \multirow{2}{*}{} & \multicolumn{4}{c|}{Input 5x5} & \multicolumn{4}{c|}{Input 9x9} & \multirow{2}{*}{} & \multicolumn{4}{c|}{Input 5x5} & \multicolumn{4}{c|}{Input 9x9}\tabularnewline
			\cline{3-10} \cline{12-19} 
			&  & Params. & FLOPs(G) & Time & PSNR & Params. & FLOPs(G) & Time & PSNR &  & Params. & FLOPs(G) & Time & PSNR & Params. & FLOPs(G) & Time & PSNR\tabularnewline
			\hline 
			EDSR & \multirow{8}{*}{$2\times$} & 38.62M & 39.56{*}25 & 9.3 & 36.04 & 38.62M & 39.56{*}81 & 27.1 & 35.97 & \multirow{8}{*}{$4\times$} & 38.89M & 40.66{*}25 & 3.5 & 30.20 & 38.89M & 40.66{*}81 & 11.2 & 30.16\tabularnewline
			\cline{1-1} \cline{3-10} \cline{12-19} 
			RCAN &  & 15.31M & 15.59{*}25 & 6.7 & 36.11 & 15.31M & 15.59{*}81 & 20.4 & 36.13 &  & 15.36M & 15.65{*}25 & 2.8 & 30.27 & 15.36M & 15.65{*}81 & 7.6 & 30.33\tabularnewline
			\cline{1-1} \cline{3-10} \cline{12-19} 
			LFSSR &  & 0.81M & 25.70 & 4.2 & 37.53 & 0.81M & 83.27 & 11.3 & 37.55 &  & 1.61M & 128.44 & 6.1 & 31.23 & 1.61M & 416.16 & 16.1 & 30.83\tabularnewline
			\cline{1-1} \cline{3-10} \cline{12-19} 
			resLF &  & 6.35M & 37.06 & 7.7 & 36.57 & 6.35M & 98.71 & 20.6 & 36.64 &  & 6.79M & 39.70 & 3.4 & 30.43 & 6.79M & 50.12 & 9.7 & 30.51\tabularnewline
			\cline{1-1} \cline{3-10} \cline{12-19} 
			InterNet &  & 4.80M & 47.46 & 6.9 & 38.08 & 12.61M & 273.60 & 18.3 & 37.19 &  & 5.23M & 50.10 & 3.3 & 31.59 & 13.06M & 347.11 & 9.2 & 30.87\tabularnewline
			\cline{1-1} \cline{3-10} \cline{12-19} 
			ATO &  & 1.22M & 27.24{*}25 & 10.5 & 38.09 & 2.06M & 130.78{*}81 & 29.5 & 38.31 &  & 1.36M & 28.08{*}25 & 6.8 & 31.25 & 3.01M & 132.45{*}81 & 20.3 & 31.40\tabularnewline
			\cline{1-1} \cline{3-10} \cline{12-19} 
			LF-DFnet &  & 3.94M & 57.22 & 6.4 & 38.42 & 22.16M & 200.46 & 17.6 & 38.40 &  & 3.99M & 57.31 & 3.2 & 31.86 & 22.21M & 204.56 & 8.9 & 31.68\tabularnewline
			\cline{1-1} \cline{3-10} \cline{12-19} 
			Ours &  & 0.48M & 13.03 & 1.5 & 38.74 & 1.36M & 63.63 & 4.3 & 38.85 &  & 0.51M & 18.24 & 0.9 & 32.07 & 1.39M & 80.50 & 2.1 & 32.26\tabularnewline
			\hline 
	\end{tabular}}
	\label{table_params}
\end{table*}

\begin{table*}
	\centering
	\renewcommand\arraystretch{1.4}
	\caption{Comparisons of our network with different number of attention modules. The best results are in \textcolor{red}{red} and the second best results are in \textcolor{blue}{blue}.}
	\resizebox{\textwidth}{45pt}{
		\begin{tabular}{|c|c|c|c|c|c|c|c|c|c|ccccc|c|}
			\hline 
			\multirow{2}{*}{Method} & \multirow{2}{*}{Scale} & \multirow{2}{*}{Params.} & \multicolumn{5}{c|}{Dataset (Input 5x5)} & \multirow{2}{*}{\textbf{Avg}} & \multirow{2}{*}{Params.} & \multicolumn{5}{c|}{Dataset (Input 9x9)} & \multirow{2}{*}{\textbf{Avg}}\tabularnewline
			\cline{4-8} \cline{11-15} 
			&  &  & EPFL & HCInew & HCIold & INRIA & STFgantry &  &  & \multicolumn{1}{c|}{EPFL} & \multicolumn{1}{c|}{HCInew} & \multicolumn{1}{c|}{HCIold} & \multicolumn{1}{c|}{INRIA} & STFgantry & \tabularnewline
			\hline 
			DDAN with 2ABs & \multirow{4}{*}{$2\times$} & 0.35M & 34.89/0.9770 & 37.52/0.9791 & 44.31/0.9941 & 37.20/0.9851 & 38.97/0.9906 & 38.58/0.9852 & 0.78M & 35.18/0.9772 & 36.85/0.9762 & 44.11/0.9928 & 37.62/0.9836 & 38.60/0.9911 & 38.47/0.9842\tabularnewline
			\cline{1-1} 
			DDAN with 3ABs &  & 0.41M & 34.99/\textcolor{red}{0.9774} & 37.60/0.9795 & 44.38/0.9944 & 37.28/0.9853 & 39.06/0.9910 & 38.69/0.9855 & 1.06M & 35.36/0.9780 & 37.09/0.9771 & 44.30/0.9943 & 37.86/0.9848 & 38.73/0.9919 & 38.69/0.9852\tabularnewline
			\cline{1-1} 
			DDAN with 4ABs &  & 0.48M & \textcolor{blue}{35.04}/\textcolor{red}{0.9774} & \textcolor{red}{37.68}/\textcolor{red}{0.9799} & \textcolor{blue}{44.47}/\textcolor{red}{0.9946} & \textcolor{red}{37.34}/\textcolor{red}{0.9855} & \textcolor{blue}{39.14}/\textcolor{blue}{0.9913} & \textcolor{red}{38.74}/\textcolor{red}{0.9857} & 1.36M & \textcolor{blue}{35.48}/\textcolor{red}{0.9786} & \textcolor{blue}{37.31}/\textcolor{red}{0.9780} & \textcolor{blue}{44.41}/\textcolor{blue}{0.9944} & \textcolor{blue}{37.97}/\textcolor{red}{0.9850} & \textcolor{blue}{39.08}/\textcolor{blue}{0.9926} & \textcolor{blue}{38.85}/\textcolor{red}{0.9857}\tabularnewline
			\cline{1-1} 
			DDAN with 5ABs &  & 0.53M & \textcolor{red}{35.05}/\textcolor{red}{0.9774} & \textcolor{red}{37.68}/\textcolor{red}{0.9799} & \textcolor{red}{44.49}/\textcolor{red}{0.9946} & \textcolor{red}{37.34}/\textcolor{red}{0.9855} & \textcolor{red}{39.15}/\textcolor{red}{0.9913} & \textcolor{red}{38.74}/\textcolor{red}{0.9857} & 1.61M & \textcolor{red}{35.49}/\textcolor{red}{0.9786} & \textcolor{red}{37.32}/\textcolor{red}{0.9780} & \textcolor{red}{44.43}/\textcolor{red}{0.9945} & \textcolor{red}{37.98}/\textcolor{red}{0.9850} & \textcolor{red}{39.09}/\textcolor{red}{0.9927} & \textcolor{red}{38.86}/\textcolor{red}{0.9857}\tabularnewline
			\hline 
			DDAN with 2ABs & \multirow{4}{*}{$4\times$} & 0.39M & 29.02/0.9170 & 31.31/0.9198 & 37.12/0.9716 & 31.24/0.9506 & 30.51/0.9411 & 31.84/0.9401 & 0.82M & 29.06/0.9165 & 31.01/0.9129 & 36.24/0.9657 & 31.33/0.9476 & 30.02/0.9406 & 31.53/0.9367\tabularnewline
			\cline{1-1} 
			DDAN with 3ABs &  & 0.45M & 29.11/0.9176 & 31.49/0.9207 & 37.26/0.9721 & 31.35/0.9510 & 30.63/0.9416 & 31.97/0.9406 & 1.09M & 29.30/0.9180 & 31.22/0.9142 & 36.55/0.9676 & 31.54/0.9489 & 30.21/0.9411 & 31.76/0.9379\tabularnewline
			\cline{1-1} 
			DDAN with 4ABs &  & 0.51M & \textcolor{blue}{29.19}/\textcolor{blue}{0.9178} & \textcolor{blue}{31.60}/\textcolor{red}{0.9211} & \textcolor{blue}{37.36}/\textcolor{blue}{0.9722} & \textcolor{red}{31.44}/\textcolor{red}{0.9513} & \textcolor{blue}{30.72}/\textcolor{red}{0.9419} & \textcolor{blue}{32.07}/\textcolor{red}{0.9409} & 1.39M & \textcolor{blue}{29.75}/\textcolor{red}{0.9235} & \textcolor{blue}{31.48}/\textcolor{blue}{0.9216} & \textcolor{blue}{36.97}/\textcolor{blue}{0.9736} & \textcolor{blue}{32.05}/\textcolor{red}{0.9543} & \textcolor{blue}{31.04}/\textcolor{red}{0.9539} & \textcolor{blue}{32.26}/\textcolor{red}{0.9454}\tabularnewline
			\cline{1-1} 
			DDAN with 5ABs &  & 0.57M & \textcolor{red}{29.20}/\textcolor{red}{0.9179} & \textcolor{red}{31.63}/\textcolor{red}{0.9211} & \textcolor{red}{37.38}/\textcolor{red}{0.9723} & \textcolor{red}{31.44}/\textcolor{red}{0.9513} & \textcolor{red}{30.73}/\textcolor{red}{0.9419} & \textcolor{red}{32.08}/\textcolor{red}{0.9409} & 1.64M & \textcolor{red}{29.78}/\textcolor{red}{0.9235} & \textcolor{red}{31.51}/\textcolor{red}{0.9217} & \textcolor{red}{36.98}/\textcolor{red}{0.9736} & \textcolor{red}{32.06}/\textcolor{red}{0.9543} & \textcolor{red}{31.06}/\textcolor{red}{0.9539} & \textcolor{red}{32.28}/\textcolor{red}{0.9454}\tabularnewline
			\hline 
	\end{tabular}}
	\label{table_att_number}
\end{table*}

\textbf{Quantitative Results}. The quantitative results are listed in Table \ref{table_results}. To make a comprehensive comparison, we present the results on $5\times 5$ and $9\times 9$ LFs for $2\times$ SR and $4\times$ SR. It can be observed that our network achieves the best performance on 4 of 5 datasets (i.e. HCInew \cite{honauer2016dataset}, HCIold \cite{wanner2013datasets}, EPFL \cite{EPFL} and INRIA \cite{INRIA}). Moreover, compared to state-of-the-art LFSR methods ATO \cite{ATO} and LF-DFnet \cite{wang2020light}, our method improves the PSNR around 1 dB for $2\times$ SR, 0.6 dB for $4\times$ SR on small-disparity datasets (i.e., EPFL \cite{EPFL}, INRIA \cite{INRIA}), and around 0.3 dB for $2\times$ SR, 0.2 dB for $4\times$ SR on mid-disparity datasets (i.e., HCInew \cite{honauer2016dataset} and HCIold \cite{wanner2013datasets}), which demonstrate the effectiveness of our network to capture complementary information. However, our network performs less superior on the STFgantry \cite{STFgantry} dataset with large disparities. That is because, our network does not explicitly address the disparity issue and tries to cover the corresponding pixels by its large receptive field. As demenstrated in \cite{wang2020light}, it is difficult for network to incorporate the complementary information among different views without alignment of views when dealing with large disparities.

\begin{figure*}[t]
\vspace{-0.0cm}
\centering{
{\includegraphics[width=15cm]{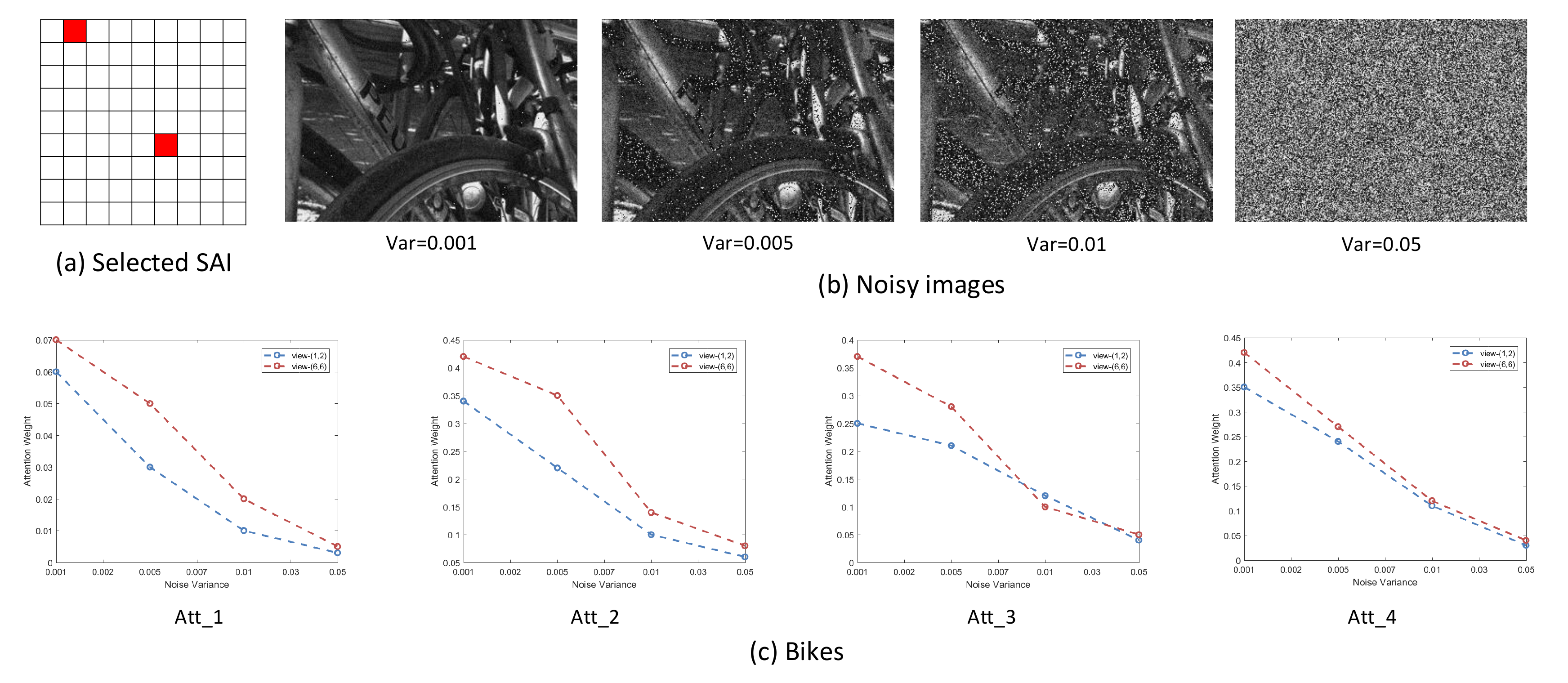}}}

\caption{Visualization of the attention weights for noisy images with different variance. (a) The two SAIs are the two perspectives that we choose to add noise to. (b) The variance of noise levels contain 0.001, 0.005, 0.01, 0.05. (c) The attention weights for the two SAIs within each attention module.}
\label{figure_att_noise}
\end{figure*}

Since LF image SR methods can super-resolve all perspectives of an LF, we further analyze the performance of these methods of each SAI on scene "Rusty Fency" from the EPFL \cite{EPFL} dataset. Due to the occlusions and specularities, the content of different SAIs is not identical, resulting in varying PSNRs among different SAIs. Since SISR methods super-resolve each SAI independently without using the complementary information, we only compare the PSNR of each SAI of learning-based LFSR methods. As shown in Fig. \ref{figure_std}, we visually compare $9\times 9$ SAIs on scene "Rusty Fency" for $2\times$ SR. It can be observed that our method achieves better performance with a relatively balanced distribution. That is because, our method can fully exploit the complementary information among different views and channels by using our dense dual-attention module.

\textbf{Qualitative Results}. The visual comparisons of different methods are shown in Figs. \ref{figure_2xSR_bicycle}, \ref{figure_2xSR_herbs} for $2\times$ SR and Figs. \ref{figure_4xSR_bedroom}, \ref{figure_4xSR_cards} for $4\times$ SR, respectively. It can be observed from the zoom-in regions that SISR methods cannot reconstruct reliable details, especially for $4\times$ SR. The scene "Bedroom" from HCInew \cite{honauer2016dataset} contains complex stripes which are hard for SR methods to recover. The results achieved by bicubic interpolation and SISR methods are blurry on the stripes regions in Fig. \ref{figure_4xSR_bedroom}. The scene "Cards" from STFgantry \cite{STFgantry} is a challenging scene due to the occlusions and complex structures. The results achieved by EDSR \cite{EDSR} show blurring artifacts around the words, and the results achieved by RCAN \cite{RCAN} contain ghosting artifacts.

\begin{figure*}[h]
	\vspace{-0.0cm}
	\centering{
		{\includegraphics[width=17cm]{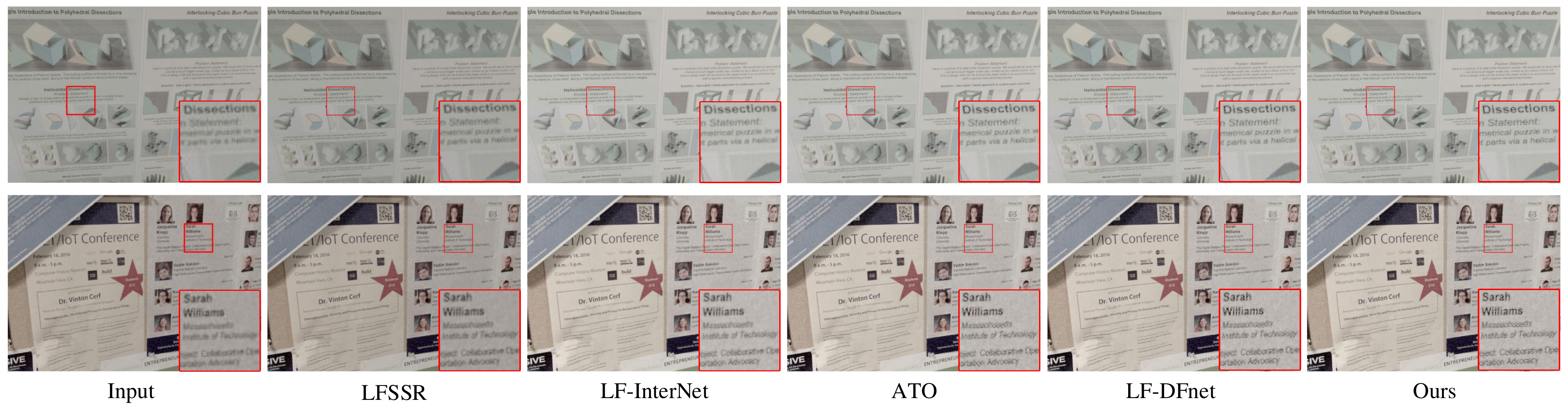}}
	}
	\caption{Visual comparisons of learning-based LFSR methods on unseen scenes for $2\times$ SR. }
	\label{figure_real_visual}
\end{figure*}

\begin{figure}[h]
	\vspace{-0.0cm}
	\centering
	\includegraphics[width=8.8cm]{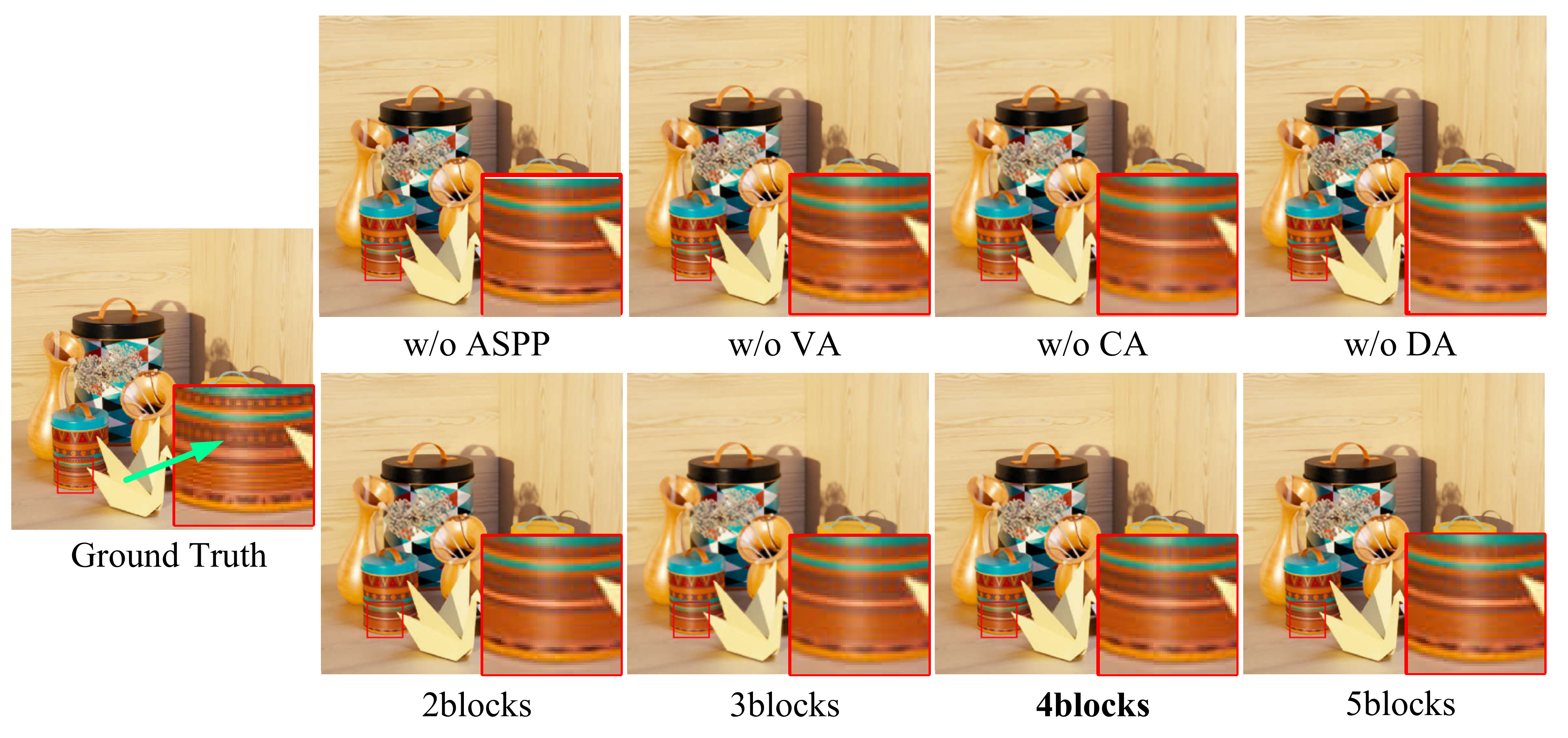}
	
	\caption{Visual results of different variants of our network on scene "Origami" for 4$\times$ SR.}
	\label{figure_ablation_results}
\end{figure}

As compared to SISR methods, LFSR methods generate better results with more high-frequency details and less artifacts. Specifically, the central images produced by our method are more close to the groundtruth images. For $2\times$ SR, as shown in Figs. \ref{figure_2xSR_bicycle} and \ref{figure_2xSR_herbs}, LF-InterNet \cite{LF-InterNet}, ATO \cite{ATO} and our method can suppress the artifacts and recover accurate textures due to the utilization of complementary information. For $4\times$ SR, as shown in Figs. \ref{figure_4xSR_bedroom} and \ref{figure_4xSR_cards}, some extent of blurring and artifacts are generated by ATO \cite{ATO} and LF-DFnet \cite{wang2020light} such as the strips on the wall in Fig. \ref{figure_4xSR_bedroom} and the letters in Fig. \ref{figure_4xSR_cards}. In contrast, our method can produce more faithful results by adaptively capture informative features across different views and channels.

\begin{table*}
\renewcommand\arraystretch{1.4}
\centering
\caption{Ablation study on the effectiveness of view attention and channel attention in our network. Average PSNR and SSIM for $2\times $ and $4\times $ SR on all datasets are reported. The best results are in \textcolor{red}{red} and the second best results are in \textcolor{blue}{blue}.}

\begin{tabular}{|c|c|c|ccccc|c|}
	\hline 
	\multirow{2}{*}{Model} & \multirow{2}{*}{Scale} & \multirow{2}{*}{Params.} & \multicolumn{5}{c|}{Dataset (Input 9x9)} & \multirow{2}{*}{\textbf{Avg}}\tabularnewline
	\cline{4-8} 
	&  &  & \multicolumn{1}{c|}{EPFL} & \multicolumn{1}{c|}{HCInew} & \multicolumn{1}{c|}{HCIold} & \multicolumn{1}{c|}{INRIA} & STFgantry & \tabularnewline
	\hline 
	DDAN w/o ASPP & \multirow{5}{*}{$2\times$} & 1.36M & \textcolor{blue}{35.24}/0.9782 & \textcolor{blue}{37.14}/0.9772 & \textcolor{blue}{44.31}/0.9942 & \textcolor{blue}{37.13}/\textcolor{blue}{0.9846} & \textcolor{blue}{38.92}/\textcolor{blue}{0.9921} & \textcolor{blue}{38.55}/\textcolor{blue}{0.9853}\tabularnewline
	\cline{1-1} \cline{3-3} 
	DDAN w/o VA &  & 1.36M & 34.91/0.9780 & 37.11/\textcolor{blue}{0.9774} & 44.13/0.9941 & 36.69/0.9842 & 38.73/0.9919 & 38.31/0.9851\tabularnewline
	\cline{1-1} \cline{3-3} 
	DDAN w/o CA &  & 1.36M & 34.97/\textcolor{blue}{0.9783} & 37.06/0.9771 & 44.27/\textcolor{blue}{0.9943} & 36.97/0.9845 & 38.69/0.9914 & 38.39/0.9851\tabularnewline
	\cline{1-1} \cline{3-3} 
	DDAN w/o DA &  & 1.36M & 34.88/0.9767 & 36.99/0.9768 & 44.13/0.9942 & 36.61/0.9843 & 38.65/0.9913 & 38.25/0.9847\tabularnewline
	\cline{1-1} \cline{3-3} 
	DDAN &  & 1.36M & \textcolor{red}{35.48}/\textcolor{red}{0.9786} & \textcolor{red}{37.31}/\textcolor{red}{0.9780} & \textcolor{red}{44.41}/\textcolor{red}{0.9944} & \textcolor{red}{37.97}/\textcolor{red}{0.9850} & \textcolor{red}{39.08}/\textcolor{red}{0.9926} & \textcolor{red}{38.85}/\textcolor{red}{0.9857}\tabularnewline
	\hline 
	DDAN w/o ASPP & \multirow{5}{*}{$4\times$} & 1.39M & 29.57/0.9213 & \textcolor{blue}{31.38}/\textcolor{blue}{0.9210} & \textcolor{blue}{36.85}/\textcolor{blue}{0.9731} & \textcolor{blue}{31.87}/\textcolor{blue}{0.9519} & \textcolor{blue}{30.89}/\textcolor{blue}{0.9518} & \textcolor{blue}{32.11}/\textcolor{blue}{0.9438}\tabularnewline
	\cline{1-1} \cline{3-3} 
	DDAN w/o VA &  & 1.39M & 29.53/0.9213 & 31.24/0.9203 & 36.71/0.9719 & 31.74/0.9511 & 30.73/0.9505 & 31.99/0.9430\tabularnewline
	\cline{1-1} \cline{3-3} 
	DDAN w/o CA &  & 1.39M & \textcolor{blue}{29.62}/\textcolor{blue}{0.9214} & 31.20/0.9204 & 36.79/0.9715 & 31.82/0.9513 & 30.69/0.9509 & 32.02/0.9431\tabularnewline
	\cline{1-1} \cline{3-3} 
	DDAN w/o DA &  & 1.39M & 29.49/0.9211 & 31.11/0.9202 & 36.64/0.9708 & 31.69/0.9509 & 30.62/0.9503 & 31.91/0.9427\tabularnewline
	\cline{1-1} \cline{3-3} 
	DDAN &  & 1.39M & \textcolor{red}{29.75}/\textcolor{red}{0.9235} & \textcolor{red}{31.48}/\textcolor{red}{0.9216} & \textcolor{red}{36.97}/\textcolor{red}{0.9736} & \textcolor{red}{32.05}/\textcolor{red}{0.9543} & \textcolor{red}{31.04}/\textcolor{red}{0.9539} & \textcolor{red}{32.26}/\textcolor{red}{0.9454}\tabularnewline
	\hline 
\end{tabular}
\label{table_ablation}
\end{table*}

\begin{table*}
	\renewcommand\arraystretch{1.4}
	\caption{Comparisons of different sturctures between cascaded structure and our dual-attention structure.}
	\centering
	\begin{tabular}{|c|c|ccccc|c|}
		
		\hline 
		\multirow{2}{*}{Model} & \multirow{2}{*}{Scale} & \multicolumn{5}{c|}{Dataset (Input 9x9)} & \multirow{2}{*}{\textbf{Avg}}\tabularnewline
		\cline{3-7} 
		&  & \multicolumn{1}{c|}{EPFL} & \multicolumn{1}{c|}{HCInew} & \multicolumn{1}{c|}{HCIold} & \multicolumn{1}{c|}{INRIA} & STFgantry & \tabularnewline
		\hline 
		VA-CA & \multirow{2}{*}{$2\times$} & 34.76/0.9771 & 36.74/0.9762 & 43.93/0.9931 & 37.26/0.9824 & 38.37/0.9902 & 38.21/0.9838\tabularnewline
		\cline{1-1} 
		DDAN &  & 35.48/0.9786 & 37.31/0.9780 & 44.41/0.9944 & 37.97/0.9850 & 39.08/0.9926 & 38.85/0.9857\tabularnewline
		\hline 
		VA-CA & \multirow{2}{*}{$4\times$} & 29.11/0.9215 & 30.67/0.9198 & 36.24/0.9713 & 31.41/0.9530 & 30.27/0.9515 & 31.54/0.9434\tabularnewline
		\cline{1-1} 
		DDAN &  & 29.75/0.9235 & 31.48/0.9216 & 36.97/0.9736 & 32.05/0.9543 & 31.04/0.9539 & 32.26/0.9454\tabularnewline
		\hline 
	\end{tabular}
	\label{table_structure}
\end{table*}

\begin{table*}
	\vspace{-0.0cm}
	\renewcommand\arraystretch{1.4}
	\centering
	\caption{Comparisons of attention blocks with different reduction ratios. The best results are in \textcolor{red}{red} and the second best results are in \textcolor{blue}{blue}.}
	\begin{tabular}{|c|c|c|c|c|c|c|c|}
		\hline 
		\multirow{2}{*}{Method} & \multirow{2}{*}{Scale} & \multicolumn{5}{c|}{Dataset (Input 5x5)} & \multirow{2}{*}{\textbf{Avg}}\tabularnewline
		\cline{3-7} 
		&  & EPFL & HCInew & HCIold & INRIA & STFgantry & \tabularnewline
		\hline 
		DDAN with ratio 2 & \multirow{4}{*}{$2\times$} & \textcolor{red}{35.04}/\textcolor{red}{0.9774} & \textcolor{red}{37.68}/\textcolor{red}{0.9799} & \textcolor{blue}{44.47}/\textcolor{red}{0.9946} & \textcolor{red}{37.34}/\textcolor{red}{0.9855} & \textcolor{red}{39.14}/\textcolor{red}{0.9913} & \textcolor{red}{38.74}/\textcolor{red}{0.9857}\tabularnewline
		\cline{1-1} 
		DDAN with ratio 4 &  & \textcolor{blue}{35.02}/\textcolor{red}{0.9774} & \textcolor{blue}{37.63}/\textcolor{red}{0.9799} & \textcolor{red}{44.48}/\textcolor{blue}{0.9945} & \textcolor{blue}{37.31}/0.9853 & \textcolor{blue}{39.10}/\textcolor{red}{0.9913} & \textcolor{blue}{38.72}/\textcolor{red}{0.9857}\tabularnewline
		\cline{1-1} 
		DDAN with ratio 8 &  & 35.01/0.9773 & 37.61/0.9798 & 44.42/0.9944 & 37.30/\textcolor{blue}{0.9854} & 39.08/0.9911 & 38.69/0.9856\tabularnewline
		\cline{1-1} 
		DDAN with ratio 16 &  & 35.01/0.9773 & 37.60/0.9797 & 44.41/0.9944 & 37.28/0.9853 & 39.06/0.9910 & 38.68/0.9856\tabularnewline
		\hline 
		DDAN with ratio 2 & \multirow{4}{*}{$4\times$} & \textcolor{red}{29.19}/\textcolor{red}{0.9178} & \textcolor{red}{31.60}/\textcolor{red}{0.9211} & \textcolor{red}{37.36}/\textcolor{red}{0.9722} & \textcolor{red}{31.44}/\textcolor{red}{0.9513} & \textcolor{red}{30.72}/\textcolor{red}{0.9419} & \textcolor{red}{32.07}/\textcolor{red}{0.9409} \tabularnewline
		\cline{1-1} 
		DDAN with ratio 4 &  & \textcolor{blue}{29.15}/\textcolor{red}{0.9178} & \textcolor{blue}{31.58}/\textcolor{blue}{0.9210} & \textcolor{blue}{37.32}/\textcolor{red}{0.9722} & \textcolor{blue}{31.41}/\textcolor{blue}{0.9512} & \textcolor{red}{30.72}/\textcolor{blue}{0.9418} & \textcolor{blue}{32.05}/\textcolor{red}{0.9409}\tabularnewline
		\cline{1-1} 
		DDAN with ratio 8 &  & 29.11/0.9177 & 31.55/0.9210 & 37.30/0.9720 & 31.39/0.9511 & 30.68/0.9417 & 32.02/0.9408\tabularnewline
		\cline{1-1} 
		DDAN with ratio 16 &  & 29.12/0.9177 & 31.51/0.9208 & 37.29/0.9721 & 31.38/0.9511 & 30.64/0.9416 & 32.01/0.9407\tabularnewline
		\hline 
	\end{tabular}
	\label{table_reduction_ratios}
	\vspace{-0.0cm}
\end{table*}

\textbf{Computational Efficiency}. To further test the performance of our method, we compare the number of network parameters, FLOPs and running time (in seconds) of our network as compared to other networks. All the methods were implemented on a computer with Intel CPU i7-8700 @ 3.70GHz, 32 GB RAM and an NVIDIA GeForce RTX 2080Ti. As shown in Table \ref{table_params}, LFSR methods achieve higher PSNR/SSIM scores and have fewer parameters/FLOPs than SISR methods. Moreover, compared with resLF \cite{resLF} and LF-InterNet \cite{LF-InterNet}, our network can achieve better performance with relatively fewer parameters. Note that, although our method achieves comparable performance with ATO \cite{ATO} and LF-DFnet \cite{wang2020light} on $5\times 5$ and $9\times 9$ LFs, the number of parameters, FLOPs and running time of our network are much fewer than those of ATO \cite{ATO} and LF-DFnet \cite{wang2020light}. The above results demonstrate the high efficiency of our proposed method.

\textbf{Generalization to Unseen Scenes}. To test the generalization ability of LFSR methods, we conduct experiments on an unseen real-world dataset named UCSD \cite{2016A}. Specifically, we directly apply LFSR methods to super-resolve LFs without downsampling. As shown in Fig. \ref{figure_real_visual}, learning-based LFSR methods can improve the LF spatial resolution on unseen scenes. It can be observed that our method produces images with sharper textures, which demonstrates that our method can be generalized to super-resolve LF images on unseen scenes.

\subsection{Ablation Study}
In this section, we conduct ablation experiments to investigate the contributions of different components in our model, including the ASPP module, the view attention module in RVAB and the channel attention module in RCAB. The visual results of our network with different components are shown in Fig. 11. It can be observed that our complete network can recover images with more clear structure and less artifacts. Moreover, we investigate the SR performance with respect to the number of attention blocks, and the reduction rations in attention blocks. Finally, we compare our dual-attention structure with cascaded structure to demonstrate its effectiveness.

\textbf{DDAN w/o ASPP module}. The ASPP module is used to extract multi-scale features. To demonstrate the effectiveness of the ASPP module, we remove this module and use residual blocks to extract shallow features. As shown in Table \ref{table_ablation}, the PSNR value is decreased from 38.85 dB to 38.55 dB for $2\times$ SR and from 32.26 dB to 32.11 dB for $4\times$ SR without using the ASPP module.

\textbf{DDAN w/o VA}. View attention branch is used to capture discriminative features across different views. To validate the effectiveness of the view attention module in RVAB, we first visualized the attention weights for two different LF images in Fig. \ref{figure_att_weights}. We compare each attention module in the view attention branch. It can be seen that the attention weights are content-aware and are different for different LF images.

To further investigate how the input content affect the attention weights, we added additive Gaussian noise with various noise levels in the selected two SAIs (located at $(1, 2)$ and $(6, 6)$). As shown in Fig. \ref{figure_att_noise}, the attention weights decrease as the noise level increases, which is consistent with our intuition. That is, the noisy images contribute less information for SR. Moreover, even in the case of high noise level (variance equals to 0.05), the attention weight is not equal to 0. That is because, the feature extraction and dense connection in branch reduce the direct influence of the noisy images on the attention weights.

Finally, we removed the view attention module in all RVABs and retrained our network with comparable number of parameters. It can be observed from Table \ref{table_ablation} that the PSNR value is decreased from 38.85 dB to 38.31 dB for $2\times$ SR and from 32.26 dB to 31.99 dB for $4\times$ SR. The SSIM value is also decreased for both $2\times$ SR and $4\times$ SR. That is because, only residual blocks in RVAB cannot effectively incorporate informative features from different views, which hinders our network from exploiting discriminative view-wise information.

\textbf{DDAN w/o CA}. We removed the channel attention module in all RCABs and retrained our network to investigate their contributions. It can be observed from Table \ref{table_ablation} that the PSNR value is decreased from 38.85 dB to 38.39 dB for $2\times$ SR and from 32.26 dB to 32.02 dB for $4\times$ SR. The SSIM value is also decreased both for $2\times$ SR and for $4\times$ SR. That is because, channel attention branch is used to model the interdependencies across different channels and only using residual blocks in RCAB cannot select important features from various channels, which leads to the performance degradation.

\textbf{DDAN w/o DA}. We investigated the benefit of the dual-attention module by removing the view attention and channel attention simultaneously. That is, we only use residual blocks in view attention branch and channel attention branch. It can be observed from Table \ref{table_ablation} that both PSNR and SSIM are decreased because treating all views and channels equally limits the representational ability of our network.

\textbf{Number of ABs}. Table \ref{table_att_number} shows the reconstruction accuracy with respect to the number of attention blocks (ABs) in our network. It can be observed that the accuracy consistently improves as the number of attention blocks increases. However, the improvements tend to be saturated when the number of ABs is larger than 4. Specifically, when the number of attention blocks is increased from 4 to 5, the PSNR improvement is less than 0.03 dB. Fig. \ref{figure_ablation_results} provides the visual comparisons achieved by our network with different number of attention blocks on scene "Origami". As the number of attention blocks increases, our network can produce better quality HR images. Since the number of network parameters grows as the number of attention blocks increases, we decided to use 4 attention blocks in both view attention and channel attention branch to achieve a good tradeoff between reconstruction accuracy and computational efficiency.

\textbf{Reduction Ratios.} In order to investigate the performance of attention blocks with different reduction ratios, we conduct experiments with four reduction ratios (i.e., $2,4,8,16$). It can be observed from Table \ref{table_reduction_ratios} that our network with the reduction ratios being set to 2 achieves relatively good performance. Therefore, the reduction ratios in both RVAB and RCAB are set to 2.

\textbf{Dual-branch Attention Structure}. We compare our dual-attention structure with cascaded structure (VA-CA) to demonstrate its effectiveness, as shown in Table \ref{table_structure}. In the cascaded structure, we perform view attention and channel attention in one branch. For fair comparison, the same number of attention blocks are used. It can be observed that our dual-attention structure can achieve better performance. That is because, both view-wise and channel-wise information are beneficial to the SR performance. By using our dual-branch structure, the view-wise and channel-wise information can be decoupled into two separate branches, and our proposed attention blocks can better learn the view-wise and channel-wise statistics within each branch.

\section{Conclusion and Future Work}
In this paper, we propose a dense dual-attention network (DDAN) for light field image super-resolution. To make fully use of view-specific characteristics among SAIs, we introduce a view attention branch by stacking RCABs with densely connected structure. Moreover, we develop a channel attention branch to selectively focus on discriminative information across all channels at the same view. Experimental results on public datasets demonstrate the effectiveness of our method. Our network achieves superior SR performance with relatively fewer parameters and a small computational cost, as compared to state-of-the-art single image and LF image SR methods.

In the future, we will investigate generative adversarial network (GAN) to generate realistic textures for LF image SR. Moreover, we will try to develop an effective method to fully take advantage of the complementary information in LF images with large disparities.

\bibliographystyle{IEEEtran}
\bibliography{DDAN}

\begin{IEEEbiography}[{\includegraphics[width=1in,height=1.25in,clip,keepaspectratio]{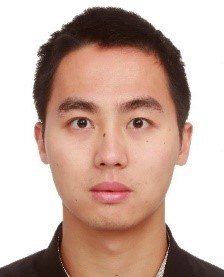}}]{Yu Mo} received the B.E. degree in information and communication engineering from University of Electronic Science and Technology of China, Chengdu, China, in 2014, and the M.E. degree with the College of Electronic Science, National University of Defense Technology (NUDT), Changsha, China, in 2016, where he is currently working toward the Ph.D. degree. His research interests include light field super-resolution and depth estimation.
\end{IEEEbiography}
\vspace{-20 mm}
\begin{IEEEbiography}[{\includegraphics[width=1in,height=1.25in,clip,keepaspectratio]{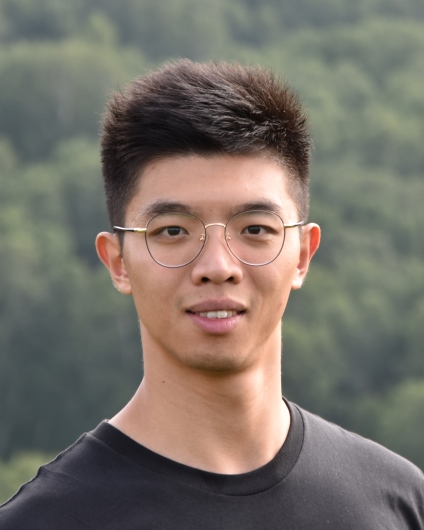}}]{Yingqian Wang} received the B.E. degree in electrical engineering from Shandong University (SDU), Jinan, China, in 2016, and the M.E. degree in information and communication engineering from National University of Defense Technology (NUDT), Changsha, China, in 2018. He is currently pursuing the Ph.D. degree with the College of Electronic Science and Technology, NUDT. He has authored several papers in journals and conferences such as TPAMI, TIP, CVPR and ECCV. His research interests focus on low-level vision, particularly on light field imaging and image super-resolution.
\end{IEEEbiography}
\vspace{-20 mm}
\begin{IEEEbiography}[{\includegraphics[width=1in,height=1.25in,clip,keepaspectratio]{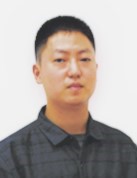}}]{Chao Xiao}received the B.E. degree in Communication Engineering and the M.E. degree in Information and Communication Engineering from the National University of Defense Technology (NUDT), Changsha, China, in 2016 and 2018, respectively. He is currently pursuing the Ph.D. degree with the College of Electronic Science and Technology, NUDT. His research interests include light field imaging, camera calibration and infrared small target detection..
\end{IEEEbiography}
\vspace{-20 mm}
\begin{IEEEbiography}[{\includegraphics[width=1in,height=1.25in,clip,keepaspectratio]{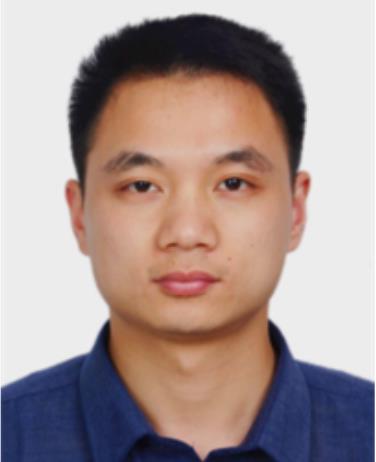}}]{Jungang Yang} received the B.E. and Ph.D. degrees from National University of Defense Technology (NUDT), in 2007 and 2013 respectively. He was a visiting Ph.D. student with the University of Edinburgh, Edinburgh from 2011 to 2012. He is currently an associate professor with the College of Electronic Science, NUDT. His research  interests include computational  imaging, image processing, compressive sensing and sparse representation. Dr. Yang received the New Scholar Award of Chinese Ministry of Education in 2012, the Youth Innovation Award and the Youth Outstanding Talent of NUDT in 2016.
\end{IEEEbiography}
\vspace{-20 mm}
\begin{IEEEbiography}[{\includegraphics[width=1in,height=1.25in,clip,keepaspectratio]{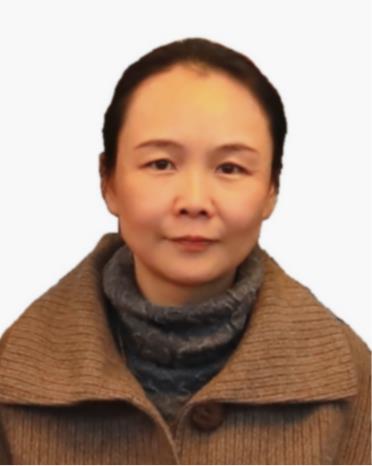}}]{Wei An} received the Ph.D. degree from the National University of Defense Technology (NUDT), Changsha, China, in 1999. She was a Senior Visiting Scholar with the University of Southampton, Southampton, U.K., in 2016. She is currently a Professor with the College of Electronic Science and Technology, NUDT. She has authored or co-authored over 100 journal and conference publications. Her current research interests include signal processing and image processing.
\end{IEEEbiography}

\end{document}